\documentclass[preprint,12pt]{aastex}

\newcommand{\mufit}{4.18}  
\newcommand{\muerror}{0.008}  
\newcommand{\siglfit}{0.36}  
\newcommand{\siglerror}{0.006}  
\newcommand{\sigrfit}{0.76}  

\newcommand{\alphafit}{0.52}
\newcommand{\alphaerror}{0.04}
\newcommand{\betafit}{12.0}
\newcommand{\betaerror}{0.31}
\newcommand{\gammafit}{0.76}
\newcommand{\gammaerror}{0.04}

\shorttitle{F Turnoff Distribution}
\shortauthors{Newby et al.}

\begin{document}

\title{F Turnoff Distribution in the Galactic Halo Using Globular Clusters as Proxies}

\author{
Matthew Newby\altaffilmark{\ref{RPI}},
Heidi Jo Newberg\altaffilmark{\ref{RPI}},
Jacob Simones\altaffilmark{ \ref{RPI}, \ref{UMN}}, 
Nathan Cole\altaffilmark{\ref{RPI}},
Matthew Monaco\altaffilmark{\ref{RPI}}
}

\altaffiltext{1}{Dept. of Physics, Applied Physics and Astronomy, Rensselaer
Polytechnic Institute Troy, NY 12180; newbym2@rpi.edu, heidi@rpi.edu\label{RPI}}

\altaffiltext{2}{School of Physics and Astronomy, University of Minnesota, Minneapolis, MN 55455\label{UMN}}

\begin{abstract}

F turnoff stars are important tools for studying Galactic halo substructure because they are plentiful, 
luminous, and can be easily selected by their photometric colors from large surveys such as the Sloan 
Digital Sky Survey (SDSS).  We describe the absolute magnitude distribution of color-selected F turnoff 
stars, as measured from SDSS data, for eleven globular clusters in the Milky Way halo.  We find that 
the $M_g$ distribution of turnoff stars is intrinsically the same for all clusters studied, 
and is well fit by two half Gaussian functions, centered at $\mu=$ \mufit, with a bright-side $\sigma=$ 
\siglfit, and with a faint-side $\sigma=$ \sigrfit.  However, the color errors and detection efficiencies 
cause the \emph{observed} $\sigma$ of the faint-side Gaussian to change with magnitude due to contamination 
from redder main sequence stars (40\% at 21st magnitude).  We present a function that will correct for 
this magnitude-dependent change in selected stellar populations, when calculating stellar density from 
color-selected turnoff stars.  We also present a consistent set of distances, ages and metallicities for 
eleven clusters in the SDSS Data Release 7. We calculate a linear correction function to Padova isochrones 
so that they are consistent with SDSS globular cluster data from previous papers.  We show that our cluster 
population falls along the Milky Way Age-Metallicity Relationship (AMR), and further find that isochrones 
for stellar populations on the AMR have very similar turnoffs; increasing metallicity and decreasing age 
conspire to produce similar turnoff magnitudes and colors for all old clusters that lie on the AMR.

\end{abstract}


\section{Introduction\label{intro}}

During the past decade, Galactic stars from the Sloan Digital Sky Survey (SDSS) have been used not 
only to discover substructure, but also to trace the density of the smooth components of the Galaxy: 
the disks and spheroid.  Because the sample of stars is large and the colors of stars are well measured, 
it has been possible to use the statistical distributions of star magnitudes to infer the distances to 
Milky Way substructure, without the need to know the absolute magnitude of every star.  In an early 
application of this technique \citep{nyetal02}, an estimate of the most common absolute magnitude of 
very blue F turnoff stars in the Sagittarius dwarf spheroidal galaxy was compared with the most common 
absolute magnitude of similar stars in the tidal debris stream to determine the distance to the tidal 
stream.  Even though there is a two magnitude spread in the absolute magnitudes of stars near the 
turnoff, these stars were used as distance indicators to discover stellar substructure and to estimate 
the distance to it.

Later studies such as \citet{betal06, juretal08} used photometric parallax of a larger color selection 
of the stars in the sample, assuming they were overwhelmingly main sequence stars, to measure the shape 
of the major components of the Milky Way or to discover substructures.  Some searches for stellar halo 
substructure \citep{retal02, gd06a, gd06b, gj06, g06, g09} have used a matched filter technique to
discover stars that have been tidally stripped from dwarf galaxies and globular clusters.

Since the \citet{nyetal02} paper, Newberg \& Yanny have continued to use the apparent magnitude of the 
turnoff stars in a discovered Milky Way substructure to estimate the distances to discovered overdensities 
of stars \citep{ynetal03,fgnl04, ny06, nycbrsw07, wnzyb09, ynetal09, nwyx10}. Sometimes the assumed 
absolute magnitude for F turnoff stars is the same as was calculated for the Sagittarius dwarf galaxy, 
and sometimes it is adjusted slightly based on available information about the stellar population of the 
substructure being studied.  In general, the distance to the structure under study is estimated from the 
peak of the apparent magnitude distribution of the bluest main sequence stars.

This technique was improved in \citet{cole08}, in which a maximum likelihood technique for measuring the 
spatial distribution of tidal debris from the Sagittarius dwarf galaxy was described.  In this paper, the 
distribution of magnitudes for F turnoff stars with $0.1<(g-r)_0<0.3$ was represented by a Gaussian with 
center $M_{g_0}=4.2$ and dispersion $\sigma_{g_0}=0.6$.  QSOs were eliminated from the sample with $(u-g)_0>0.4$.
This distribution is approximately the same as the distribution of similarly selected F turnoff stars for 
three globular clusters measured by \citet{ny06}, except that \citet{ny06} fit a slightly larger tail 
on the faint side of the peak than the bright side of the peak.  With the maximum likelihood technique,
the star number counts expected for a particular stream and smooth halo model (assuming a single 
absolute magnitude value for all turnoff stars) were broadened in the $g_0$ direction by the absolute 
magnitude distribution for the turnoff stars, before the number counts were compared with the data.  Using 
this method, one could estimate the depth of the tidal debris stream along the line of sight, and could make 
a more accurate fit to the smooth distribution of stars in the stellar halo.

Implicit in all of these studies has been the assumption that the absolute magnitude distribution of F 
turnoff stars is well represented by a single Gaussian distribution with a fixed peak and width.  One 
might expect that the actual distribution might be a function of age and metallicity of the stellar 
component.  In fact, the distribution might be different as a function of distance along a tidal debris 
stream, and different across the stellar halo and disks.  To effectively use F turnoff stars as standard 
candles, one should study the variation in the distribution of F turnoff star absolute magnitudes as a 
function of stellar population.

In this paper, we analyze eleven globular clusters in the Galactic halo, and find that the peak of the 
distribution of absolute magnitudes of F turnoff stars is typically $M_g=$ \mufit, and that the asymmetric 
distribution can be approximated by a half Gaussian on the bright side with a width of $\sigma=$ \siglfit , 
and a half Gaussian on the faint side with a width of $\sigma=$ \sigrfit.  This distribution is surprisingly 
similar for all of the globular clusters studied, which range in age from 9.5 to 13.5 Gyr and from [Fe/H]$=-1.17$
to [Fe/H]$=-2.30$.  We explain this surprising result by showing that it is consistent with the age-metallicity 
relationship for Galactic stars.  Older clusters should have fainter, redder turnoffs.  However, older
clusters also contain fewer metals, which pushes the turnoff brighter and bluer.  Although the color of 
the turnoff varies slightly from cluster to cluster, the absolute magnitude of the turnoff only shifts about 
0.1 magnitudes from the mean.

Although we find the turnoff magnitude to be similar for the clusters studied, observational effects may 
considerably change the properties of turnoff star distributions.  As one samples stars to the limit of 
the survey, the photometric errors increase.  Although these photometric errors are small compared to 
the uncertainty with which we know the absolute magnitude of a single F turnoff star, they can be large 
compared to the width of our color selection box.  For bright magnitudes, the actual colors of most of 
the stars selected are actually within the color range selected.  For magnitudes near the survey limit, 
some stars that should be selected are randomly measured with colors that are too red or too blue, and 
a larger number of stars that are too red or too blue are randomly scattered into our selection range.  
The largest flux into and out of the color selection range is on the red side of the range, and primarily 
broadens the measured width on the faint side of the peak absolute magnitude as one approaches the survey 
limit.

The results of this paper will make it possible to make more accurate measurements of the intrinsic 
density of F turnoff stars in the Milky Way's stellar halo.  We present a function that can be used 
to estimate the observed width on the faint side of the absolute magnitude distribution.  We also provide 
numbers for the fractional increase in the number counts as stars leak into the color box at distances 
where G main sequence stars are near the survey limit; and the fractional decrease in the number counts 
as F turnoff stars reach the survey limit and the large color errors scatter them out of the color 
selection range.

\section{Globular Cluster Data Selection}

Our analysis used photometric data from eleven globular clusters, taken from the SDSS database's seventh 
data release (DR7; Abazajian et al. 2009).  There are a total of seventeen globular clusters within the 
SDSS DR7 footprint, but five clusters (NGC 2419, NGC 7006, Pal3, Pal 4, and Pal 14) were eliminated  
because they are too distant, and one (Pal 1) was eliminated because there were too few stars to obtain 
accurate measures of the F turnoff star distribution using SDSS data.  A list of all clusters studied in 
this investigation can be found in Table 1.  

To determine the limits in right ascension and declination for selecting stars for each field, we used 
the SDSS SkyServer's Navigate tool.\footnote{Located at http://cas.sdss.org/dr7/en/tools/chart/navi.asp}  
The image field of view was expanded so that the cluster was clearly visible, then expanded further until 
a sizable background distribution was also contained within this view.  In general, the entire field of 
view spanned a rectangle with sides of lengths between 6 and 8 times the apparent radius of the cluster.  
The bounds of this rectangle were then used as the right ascension and declination limits in the Casjobs 
data query.

Within these limits, we selected all objects classified as STAR, and that had $(u - g)_0 > 0.4$.  The 
latter requirement was designed to avoid possible contamination from quasars \citep{ynetal00}. By 
selecting from the database of ``STAR"s, we ensured that we obtained only one instance of each object.
We extracted the extinction-corrected (denoted by the subscript `0') point-spread-function 
(PSF) apparent magnitudes with errors.  We determined $r_{clus}$, the radius within which the majority 
of the stars belong to the globular cluster, and $r_{cut}$, the radius outside of which there is little 
contamination from cluster stars, by visual inspection of the data.  Stars with $r_{clus} < r < r_{cut}$ 
were removed from the data set.  An example is shown in Figure 1.  The $r_{clus}$ and $r_{cut}$ values 
used for each cluster are included in Table 1.

The following clusters are listed in Table 1, but were excluded in some of our analyses:

NGC 5053 has a very low Zinn \& West metallicity ($[Fe/H]_{ZW84}$ $\sim$-2.58) which falls outside 
the range of the \citet{cg97} conversion scale, and is also below the minimum 
metallicity value for which Padova isochrones can be generated.  In the newer work of \citet{cbg09}, 
an $[Fe/H]$ of -2.30 is found, which is within the range of the Padova isochrones \citep{g2000, mgbg08}, 
and therefore we use this metallicity value for our analysis.  To indicate status as a potential 
outlier, results for NGC 5053 are plotted using a red-dotted series in figures showing turnoff stars 
properties.

M15 (NGC 7078) is a `core-collapsed' cluster, and so has a very compact core.  It may also have 
different dynamics and stellar distributions than standard globular clusters \citep{haur10}.
The SDSS photometric pipeline does not attempt to deblend crowded star fields,
and so information on M15 in SDSS is highly biased towards stars found on the edges of the cluster.
We do not expect this cluster to necessarily be consistent with the other clusters, but we include it 
in analysis anyway.  We indicate M15's status as potential outlier in figures showing turnoff star 
properties by plotting it with a red dotted series.

NGC 4147 contains only slightly more stars than M92, (583 versus 334) but they are more concentrated
around the turnoff due to a lack of stars below $M_g=6.5$. This lack of stars is due
to SDSS crowded-field photometry detection efficiency problems at fainter magnitudes (see Section 4).
We attempt to analyze NGC 4147 with other clusters later in our analysis, but we expect errors to be
large due to the low number of data points available.  In figures showing turnoff star properties, a 
blue dotted series is used for NGC 4147 to indicate it's low star counts and high expected errors.

Cluster M92 (NGC 6341) fell on the edge of the SDSS DR7 footprint, and relatively few stars were observed 
in the cluster.  The number of cluster stars was large enough to produce a fiducial fit with large bins in 
magnitude (0.5 magnitudes).  Therefore we were able to fit a modified Padova isochrone to this cluster.  
However, as there are so few M92 stars in our data, especially close to the turnoff, that M92 was omitted 
from the turnoff analysis.

\section{Isochrone Fitting to Determine Cluster Distances}

In order to convert our observed apparent magnitudes to absolute magnitudes, we rely on the measured
distances to each globular cluster.  Since we would like to study the effects of age and metallicity
on the absolute magnitudes of the turnoff stars, we would like measurements of these quantities that
are as accurate and as uniform as possible for our sample of globular clusters.  In this section, we 
assemble the spectroscopically determined metallicities ($[Fe/H]$) from a single group of authors;
measurements were obtained from Zinn \& West (1984) and then converted to more modern Carretta \& Gratton 
scale using the conversion provided in \citep{cg97}.  Using these metallicities, we then fit ages and 
distances to the clusters in a consistent fashion using Padova isochrones.

The Padova theoretical isochrones were fit to fiducial sequences determined from data for eleven clusters 
found in SDSS.  Stars in each cluster were separated into $g_0$ bins, then the mean and standard deviation 
($\sigma_{g-r}$) of the $(g-r)_0$ distribution in each bin was found.  Any stars in the $g_0$ bin with a 
$(g-r)_0$ value beyond 2$\sigma_{g-r}$ from the mean were rejected, and then the $(g-r)_0$ mean and $\sigma_{g-r}$ 
of the remaining population was found.  The $(g-r)_0$ mean and average $M_g$ value were accepted as a point 
on the fiducial sequence once the entire bin was within 2$\sigma_{g-r}$ of the current mean.  Isochrones 
were then fit to the fiducial sequences using distance and age as free parameters, while metallicity was 
held constant at the spectroscopically determined value.  

In our initial attempts to use this technique to determine cluster properties, we found that Padova 
isochrones that are good fits to both the main sequence and the subgiant branch require unreasonably 
high ages ($>$15 Gyr) for most clusters. The lack of agreement between theoretical isochrones and 
cluster data has been explored by previous authors.  Using eclipsing binary stars in the Hyades open 
cluster, \citet{petal03} showed that theoretical isochrones do not match true star populations; there 
are discrepancies in mass, luminosity, temperature, and radius.  In the second paper in the 
series, \citet{petal04} uses \emph{Hipparcos} parallax data for the Hyades cluster to further calibrate 
theoretical isochrones, finding that offsets in color indexes are sufficient to bring a theoretical 
isochrone in line with real main sequence data.  In the fourth and final paper of the series 
\citet{An2007b} fits isochrones to Galactic open clusters using color corrections in $(B-V)_0$ as a 
function of $M_v$, while using Cepheid variables as calibration points.  In \citet{an2009}, updated 
Yale Rotating Evolutionary Code with MARCS model atmospheres were used to produce \emph{ugriz} isochrones 
which were fit to main sequences of five globular clusters, producing ages and distances to these clusters.

Using similar techniques, we seek to calibrate Padova \emph{ugriz} isochrones to the \citet{an2009} results.  
Comparing a Padova isochrone generated from the \citet{an2009} derived age and distance for globular cluster 
NGC 6205 with our derived fiducial fit, we find that the difference between the theoretical isochrone and 
the data is very nearly linear along the main sequence (Figure 2).  Therefore, we apply a linear color 
correction function in $(g-r)_0$, holding $M_g$ as the independent variable, to the Padova isochrone to bring 
it into agreement with our fiducial fit.  We fit only to the main sequence and subgiant branch, and ignored 
the giant branch ($\sim M_g > 3.5$), as the linear trend was no longer valid brighter than the subgiant branch.  
We also did not fit to the lower main sequence, as our data does not extend to fainter absolute magnitudes.  
Our color-correction functional fit is given by the following equation:

\begin{equation}
  \Delta(g-r) = -0.015*M_g + 0.089
\label{isofunc}
\end{equation}

This function was applied to all of the colors in the model isochrones and then these models were fit 
to our fiducial sequences derived from the data.  For each cluster, we first chose a Padova isochrone 
that, modified by Equation 1, appeared to fit the fiducial sequence well.  We then chose additional
isochrones, identical except for age, which was varied about our ``by eye" best fit in increments of
0.2 Gyrs.  A Gaussian function was fit to the residuals of these isochrones fits, and the mean of this 
Gaussian was taken as the best fit age.

Errors in our fit ages were determined through a Hessian matrix method by comparing the residuals in 
the isochrone fit.  In the limit of a single fit parameter (age), the Hessian error method reduces to:

\begin{equation}
  \sigma = \left( \frac{R^2(t+2h)+R^2(t-2h) - 2R^2(t)}{8h^2} \right )^{-\frac{1}{2}}
\label{age_error}
\end{equation}

Where $R^2(t)$ is the residual of the isochrone fit at age $t$, and $h=0.2$ is the step size in the 
age determination method.  Using Equation 2, we were able to determine the age fit errors for each
cluster through the use of three isochrones:  the isochrone of best fit age, and two isochrones 
generated at the best fit age $\pm 0.4$ Gyr.

Padova isochrones fit using our correction function produce a consistent set of metallicities, ages,
and distances to our globular cluster sample, as presented in Table 1.  Cluster color-magnitude diagrams, 
fiducial fits, and modified isochrone fits are shown in Figure 3.

We compare our distance fits to distances in three other sources (\citet{DA2005}, \citet{hal97} and 
\citet{D2010}), and compare our ages to other isochrone-derived ages (\citet{DA2005}, \citet{MF2009}, 
and \citet{D2010}), in Figure 4. Our distances appear to be in excellent agreement with 
other sources.  Our ages agree to within the formal errors for each cluster, but appear to have a 
small linear systematic offset.  The ages are a very close match around 13 Gyr, but are a Gyr or 
two higher for ages of ~10 Gyr, so our age scale is slightly more compressed than the ages in 
the comparison sources.  

\section{Detection Efficiency for Stars in SDSS Globular Clusters}

It is clear from Figure 3 that the cluster data is incomplete at fainter absolute magnitudes, 
especially amongst the farther clusters: Pal 5, and NGC 4147, NGC 5024 and NGC 5053.  This 
incompleteness begins at a brighter magnitude than is expected from the SDSS detection efficiency 
for stars \citep{nyetal02}.  The poorer detection efficiency in globular clusters is due to 
difficulty in detecting faint sources in highly crowded star fields, and in particular the poor
performance of the SDSS photometric pipeline in this regime \citep{a-m08}.  For sufficiently 
crowded fields, the code cannot deblend and resolve faint stars, since they are washed out 
by much brighter stars.

To quantify the cluster detection efficiency, we examined nearby clusters with relatively complete
 CMDs and compared them to the farther, incomplete CMDs under the assumption that there are similar 
absolute magnitude distributions for the stars in nearby and distant clusters.  We select stars from 
the CMDs in a rectangular box with bounds $3.567 < M_{g_0} < 7.567$, $0.0 < (g-r)_0 < 0.8$, and a 
color cut of $(u-g)_0 < 0.4$, to coincide with areas of completeness in the near clusters (specifically 
$18.0 < g_0 < 22.0$ for NGC 6205) that are incomplete for the more distant clusters.  We choose three 
nearby clusters, NGC 6205, NGC 5904, and NGC 5272, and found that their normalized $M_{g_0}$ histograms 
in our box are, in fact, similar.  Therefore, we average them to create a reference absolute magnitude 
distribution for cluster stars.  We assume that this histogram represents the true magnitude distribution 
of globular clusters, then compare the normalized histograms of the four most distant clusters 
(NGC 4147, NGC 5024, NGC 5053, and Pal 5) to this reference.  To make the normalizations comparable 
across the incompleteness of the farther clusters, we took the average difference of the first five 
bins (which showed a similar rising trend in all of our clusters) and scaled the entire histogram
 by this value, thereby scaling the clusters by matching initial trends.  

To quantify the detection efficiency in our globular clusters, we fit the ratios between the reference 
histogram and the distant cluster histograms using a parabolic function.  We expect 
these globular clusters to have the same intrinsic absolute magnitude distribution as the bright 
clusters, but are missing faint stars that were lost to the crowded-field photometry.  The histogram 
residuals and functional fit can be seen in Figure 5, and the resulting function is given empirically as:

\begin{equation}
\label{cde}
CDE(g_0) = 
\left\{ \begin{array}{ccl}
1.0 \mbox{ if } g_0 < 20.92\\ 
0.0 \mbox{ if } g_0 > 23.57\\
1.0 - 0.14 \cdot (g_0 - 20.92)^2 \mbox{ otherwise}
\end{array} \right.
\end{equation}

We can use this function to reconstruct incomplete cluster star distributions, and to model the 
effect of SDSS crowded-field photometry at farther distances.  Errors in the parameter fits in 
Equation 3 are 0.14 $\pm$ 0.001, 20.92 $\pm$ 0.007.

\section{Absolute Magnitude Distribution of F Turnoff Stars}

\subsection{Fitting the Turnoff}

We are now prepared to characterize the absolute magnitude distribution of F turnoff stars in 
old stellar populations.  We select stars in the F turnoff region ($0.1 < (g-r)_0 < 0.3$), which is 
the color range used in the photometric F turnoff density searches in \citet{cole08}, and the range
originally chosen in \citet{nyetal02} to include stars redder than most blue horizontal branch stars, but 
bluer than the turnoff of Milky Way disk stars. We then build a histogram in $M_g$ from the cluster 
data, using a bin size of 0.2 magnitudes over the range $2.0 < M_g < 8.0$, which minimizes potential
contamination from non-cluster stars.  We then divide each bin by the cluster detection efficiency 
for the $g_0$ that corresponds to the bin's $M_g$, thereby rebuilding the component of each cluster 
lost due to the detection efficiency.

We subtract from this histogram the expected number of field stars in each magnitude bin, as determined 
from the sky area around each cluster, scaled to the sky area of cluster data.  If the cluster lies on 
the edge of the SDSS footprint, or where sections of the clusters were unresolved by SDSS, the areas 
with no data were excluded from the area used in scaling the background bins.  The magnitude histogram 
for the background was divided by the field stellar detection efficiency \citep{nyetal02} as a function 
of apparent magnitude.  

The top panel in Figure 6 shows the distribution of turnoff star $M_g$ absolute magnitudes 
in globular cluster NGC 5272.  We fit each SDSS cluster histogram with a `double-sided' Gaussian 
distribution, where the standard deviation is different on each side of the mean.  This choice 
provides a good fit to the data using a simple, well-known function.   We have no theoretical 
motivation behind our choice of fitting function; however, this function appears to effectively 
match the form of our data without over-determining the system. The form of our fit function is 
given by:

\begin{equation}
\label{dsgd}
\mathcal{G}(x; \mu, \sigma_l, \sigma_r, A) = A \cdot exp\left\lbrack -\frac{1}{2} \frac{(x-\mu)^2}{\sigma_i^2} \right\rbrack \\
\end{equation}

where: 

\begin{displaymath}
\sigma_i = 
\left\{\begin{array}{lr}
\sigma_l & if M_g \leq \mu \\
\sigma_r & if M_g > \mu
\end{array} \right.
\end{displaymath}

When normalized:

\begin{equation}
A = \frac{1}{\sqrt{2 \pi} (\frac{\sigma_l + \sigma_r}{2})  } 
\end{equation}

The fit parameters $A$, $\mu$, $\sigma_l$, and $\sigma_r$ are the amplitude, mean, left-side
standard deviation, and right-side standard deviations, respectively. We count all bins outside
of the range $2.0 < M_g < 8.0$ to be zero. 

We assumed Poisson counting errors when fitting this function to our data histograms.  Typically, 
the approximation of `square-root n' is used to simulate true Poisson errors, but we found that 
this over-emphasized low-count bins at the expense of the overall fit.  We instead modeled our 
Poisson errors in the fashion of Equation 10 from \citet{G86}.  At the standard 1-$\sigma$ 
confidence level, this equation simplifies to $\delta n = 1 + \sqrt{n+1}$.

A two stage process was used in fitting the parameters of the double-sided Gaussian function to 
the $M_g$ histograms.  First a Markov-Chain Monte Carlo technique was used to sweep the parameter 
space, then the best fit point was fed into a gradient descent algorithm.  We found that this 
two-stage fit allowed us to avoid local minima and find the true best fit.  A sample histogram and 
functional fit are provided for M3 (NGC 5272) on the top plot of Figure 6.  Results from the fits 
to ten clusters can be found in Table 2.  A Hessian matrix, calculated at the best fit parameters,
was used to determine the model errors in the parameter fit values.  This matrix was multiplied by 
2 to make it equivalent to the curvature matrix, then inverted, so that the diagonal elements are
equivalent to the squared parameter variances.  

To determine the turnoff color, $(g-r)_{TO}$, for a cluster, we fit a Gaussian profile to the $(g-r)_0$ 
histogram of all stars within 0.5 magnitudes of the $\mu$ parameter determined by the turnoff fits.  
The mean value from this Gaussian is taken as the turnoff value of the cluster.  This definition of 
turnoff color is then consistent with the turnoff color of theoretical isochrones (the bluest point of 
the isochrone turnoff).  Note that we do not want to choose the `bluest' (minimum $(g-r)_0$ value) point 
of our data, since that ignores the intrinsic color spread of the data, and could result in a bluer 
measurement of the turnoff with distance as color errors become larger. 

\subsection{F Turnoff Distribution Results}

We now examine our F turnoff distribution fit parameters as a function of the ages, metallicities, 
and distances to the clusters in our sample.  Keep in mind that throughout this section, we are
describing the turnoff magnitude in the SDSS $g$ filter and turnoff color in $(g-r)_0$.

In Figure 7, we find no significant correlation with cluster age and our fit parameters $\mu$, 
$\sigma_l$, $\sigma_r$, and $(g-r)_{TO}$.  Intuitively, as a cluster ages $\mu$ should move to higher 
magnitudes (fainter) and the turnoff color should move to smaller (redder) values of $(g-r)_0$.  The 
invariance of these parameters with age is evidence for a fairly uniform globular cluster sample and 
for the the Age-Metallicity Conspiracy discussed in Section 7.

When these fit parameters are plotted vs. metallicity, possible relationships emerge (Figure 8).  In 
general, when the metallicity of a theoretical isochrone is increased while holding other variables 
constant, the turnoff becomes more ``pinched" and dimmer - that is, the main sequence and subgiant 
branch move closer together in magnitude while the turnoff point becomes fainter.  The decrease of the 
$\sigma_l$ and $\sigma_r$ parameters with increasing $[Fe/H]$ indicates that ``pinching" is occurring, 
however, our $\mu$ fits \emph{decrease}, showing the opposite behavior from what is expected.  This is 
evidence that metallicity is not the dominant factor in determining turnoff brightness, and will be 
further explored in Section 7.  A reddening (increasing) of turnoff color is expected with increasing 
$[Fe/H]$, and is observed in our fits.  Therefore, metallicity is dominant over age in determining the 
turnoff color.  It is important to note that, for our cluster sample, the absolute magnitude distribution
has little dependence on metallicity.

Having explored the F turnoff distribution as a function of age and metallicity, we now turn our attention 
to the distance (Figure 9). As we do not expect position in the Galactic halo to significantly change 
the structure of a cluster, we expect all of our turnoff fit parameters to be intrinsically independent of 
the observed distance for similar clusters.  Any parameter dependence on distance is therefore actually a 
result of observational errors.  We see from Figure 9 that $\mu$ and $\sigma_l$ stay roughly constant with 
distance, and $\sigma_r$ rises first, then decreases.  The turnoff color ($(g-r)_{TO}$) appears to be 
invariant with distance.  In the next section, we study the observational errors contained within the SDSS 
database, and show that these errors explain the observed changes in the distribution of F turnoff stars
as a function of distance.

\section{Observational Effects of SDSS Errors}

\subsection{Photometric Errors in SDSS Star Colors}

Photometric errors widen the observed magnitude and color distributions of turnoff stars in globular 
clusters as plotted in color-magnitude diagrams, by an increasing amount with fainter magnitudes.
We now examine the photometric errors in the SDSS database and extrapolate the observational biases 
that result from these errors.  SDSS forgoes the conventional ``Pogson" logarithmic 
definition of magnitude in favor of the ``asinh magnitude" system described in \citet{letal99}.  
These systems are virtually identical at high signal-to-noise, but in the low signal-to-noise regime 
asinh magnitudes are well-behaved and have non-infinite errors.  The two magnitude systems do not 
differ noticeably in magnitude or magnitude error until fainter than 24th magnitude (signal-to-noise
$\sim$2.0), so our analysis will be valid in both systems.  Figure 10 shows the relationship 
between apparent magnitude and magnitude error in the $u_0$, $g_0$ and $r_0$ passbands for stars in 
the Palomar 5 selection field.  We fit an exponential function with an argument that is linear in 
apparent magnitude to the error vs. magnitude data, and find:

\begin{equation}
  \epsilon(u_0) = 0.0027 + e^{(0.80 u_0 - 19.2)} 
\label{epsu}
\end{equation}

\begin{equation}
  \epsilon(g_0) = 0.00031 + e^{(0.79 g_0 - 20.0)} 
\label{epsg}
\end{equation}

\begin{equation}
  \epsilon(r_0) = -0.000026 + e^{(0.80 r_0 - 19.7)}
\label{epsr}
\end{equation}

It is the negative constant term in the exponential that determines the magnitude at which the errors 
start to become significant.  It is evident from the functional fits and Figure 10 that, of the three 
studied passbands, errors in $u_0$ grow most quickly, while errors in $g_0$ are the smallest throughout.
Most of our good data is brighter than 23rd magnitude, where $g_0$ and $r_0$ magnitude errors are less 
than 0.3.  This will have little effect on the overall appearance of the H-R diagrams, however, magnitude
errors of this degree will have a noticeable affect on the observed colors, where the faint magnitude 
errors are on the order of the observed $(g-r)_0$ values.  

\subsection{F Turnoff Contamination}

Since F turnoff stars are selected through color cuts, we expect SDSS color errors to cause the turnoff 
to be contaminated by misidentified non-turnoff stars.  We took cluster NGC 6205 as a reference data set, 
as the relatively low distance (7.7 kpc) implies minimal photometric magnitude errors near the turnoff.  
Cluster NGC 6205 color errors do not become noticeable until $M_g=7.0$ ($g_0=21.43$); at our absolute 
magnitude limit of $M_g=8.0$ ($g_0=22.43$), the magnitude errors are close to 0.1 in $g_0$ and 0.15 in 
$r_0$, resulting in a maximum color error at $M_g=8.0$ equal to 0.18.  From Figure 3, we see that NGC 
6205 stars between $7.0 < M_g < 8.0$ are sufficiently red that even for the maximum color error, these 
stars are statistically unlikely to to be detected in the turnoff color cut.  We therefore can assume 
that NGC 6205 will be illustrative as an example of a cluster with an uncontaminated turnoff.

In order to understand how the errors affect a cluster with increasing distance, we must first define 
a process that will allow us to view a cluster as though it were observed by SDSS at a farther distance,
taking into account the increasing magnitude errors as apparent magnitudes increase.  We first choose a
new `effective distance' ($d_{\rm eff}$) to observe the cluster at, and then perform a `distance shift'
on each star for each of $u_0$, $g_0$ and $r_0$;  the magnitude of each star is increased by an amount 
equivalent to observing the cluster at $d_{\rm eff}$ instead of the original distance ($d_0$).  
The magnitude error associated with the new magnitude is then derived from the appropriate error equation 
(Equations 6, 7 and 8).  The original magnitude error is then subtracted in quadrature from the new 
magnitude error to produce the relative increase in error.  The new magnitude value is then modified by 
the relative increase in error:  we select a random value from a Gaussian distribution with a mean  
equivalent to the new magnitude, and with a standard deviation equal to the relative increase in error.
This random value produces the new shifted magnitude, equivalent to observing the star at $d_{\rm eff}$,
including the effects of SDSS magnitude errors.  

Having defined the distance shift process, we then separate the NGC 6205 cluster data into three 
color bins and examine bin cross-contamination as the distance increases.  The three color bins are: the 
primary `yellow' turnoff bin $0.1 < (g-r)_0 <0.3$, the `red' star bin ($(g-r)_0 > 0.3$) and the `blue' star 
bin ($(g-r)_0 < 0.1$).  Each of these bins were treated as separate data sets.  We then performed 
distance shifts on each bin at 1.0 kpc steps, up to a maximum $d_{\rm eff}$ of 80.0 kpc.  At each step 
we reinforced the $(u-g)_0$ color cut, then counted the number of stars that remained in their origin bin, 
and the number that had leaked into other color bins due to color errors.  We performed this process 100 
times with the NGC 6205 cluster data, and averaged the results to smooth out potential random errors.
We then repeated this process, but included the field stellar detection efficiency in our calculations
in order to represent the actual observed turnoff population.

Figure 11 illustrates the composition of the selected turnoff stars ($0.1 < (g-r)_0 < 0.3$) as a fraction 
of the original `true' turnoff count, and as a function of distance.  This calculation assumed 100\% 
detection efficiency for the shifted stars.  The results of applying the field stellar detection efficiency 
is shown by the dotted lines. The trends in this figure are color-coded by the bin of origin; the black 
trend is a sum of all lower curves, and represents the total number of stars detected as turnoff stars 
for a given distance.  Readily apparent from Figure 11 is the quick influx of `red' stars into the 
turnoff between distances of (10.0-20.0 kpc), and the constant loss of true turnoff stars with distance.  

Red G stars that lie just below the turnoff are on the densely populated main sequence, and have $(g-r)_0$ 
values just above our turnoff color cut of 0.3.  Even slight color error perturbations will tend to shift 
some these stars into the turnoff color cut, resulting in significant `red' star contamination at relatively 
low halo distances.  It is clear from Figure 11 that this is a rapid effect that occurs at fairly low distances 
(10.0-20.0 kpc), but as the errors continue to increase it becomes just as likely for a red G star to jump 
over the turnoff selection box as it is to land in it, thereby causing the red star contamination to stop
increasing around 20.0 kpc.  

It is interesting to note that the $(u-g)_0$ color cut causes some suppression of the red contamination 
effect.  Since red stars are fainter and farther down the main sequence, they have higher measured 
magnitudes and magnitude errors than true turnoff stars.  As the $u_0$ passband has the highest associated 
magnitude errors, faint stars with large errors are perturbed more in $(u-g)_0$ than in $(g-r)_0$. 
Therefore, faint red stars with large color errors may be perturbed beyond the $(u-g)_0$ cut and 
subsequently removed from the data set, even if errors would place that star in the $(g-r)_0$ cut for 
the turnoff.  The $(u-g)_0$ cut then serves to remove some of the red contamination stars from the 
turnoff selection.

As the errors increase, true F turnoff stars can only leak out of the turnoff selection box.  Since there are 
very few bluer ($(g-r)_0 < 0.1$) A-type stars or redder subgiants in globular clusters, and these in any event 
are bright, few A stars leak into the F turnoff star selection due to errors in color.   Around the distance 
that the red star contamination stops ($\sim$25.0 kpc), the number of stars in the turnoff selection box is 
comprised of roughly 60\% true F turnoff stars and 40\% redder G star contamination.  As distances increase,
the fraction of true turnoff stars in the color selection range decreases to 50\%, and the total number of
stars selected as turnoff stars decreases.  This is a significant effect that has never been accounted 
for in previous research papers.

So that future authors may compensate for these effects, we provide analytical functions for the F turnoff 
dissipation and the red star contamination.  These fits are to the 100\% detection efficiency case
(solid curves in Figure 11), representing the effect caused by color errors only.  We represent the
 F turnoff dissipation with a 4th-order polynomial function in $d_{\rm eff}$, and fit the red 
contamination with a similar function of 7th-order, with the coefficients given by $\mathbf{a_y}$ 
and $\mathbf{a_r}$ (where $\mathbf{a}= (a_0, a_1, a_2, ...)$, subscripts corresponding to order of 
term), respectively: 

\begin{equation}
  \mathbf{a_y} = (1.06, -0.031, 0.00020, 2.54\times 10^{-6}, -2.67\times 10^{-8})
\label{yellowfit}
\end{equation}

\begin{equation}
  \mathbf{a_r} = (0.016, -0.020, 0.0066, -0.00043, 1.26\times 10^{-5}, -1.92\times 10^{-7}, 1.47\times 10^{-9}, -4.54\times 10^{-12})
\label{redfit}
\end{equation}

These functions are valid in the range of $0.0 < d_{\rm eff} < 80.0$, that is, the range of the distance
shifts used in the above analysis.

\subsection{Effects of Magnitude Errors on the Turnoff Absolute Magnitude Distribution}

In order to study the effects of SDSS errors on our measured fit parameters, we performed distance
shifts (see previous section) on the $u_0$, $g_0$ and $r_0$ magnitudes of each cluster at varying 
$d_{\rm eff}$ steps, to a maximum of 44.0 kpc.  At every $d_{\rm eff}$, we applied the $(u-g)_0 < 0.4$ 
color cut to remove stars that would have been removed as if we had selected this data from the SDSS 
database.  

At each distance shift we took into account the background subtraction and detection losses.  Since 
distance shifts must be performed on a dataset of individual stars, while the correction functions must 
be applied to histogrammed data, we performed the distance shift before background subtraction and 
detection efficiency correction could occur.  In order to keep the background subtraction consistent 
with a new, effectively more distant cluster data set, we applied an equivalent distance shift to the 
background prior to binning and subtraction.  This reproduces the effect of subtracting the background 
prior to the shift.  We then divided by the cluster detection efficiency, but with the parabola center 
shifted with the cluster to the new magnitude.  If we did not shift the detection efficiency function 
before dividing, it would be applied to the wrong portion of the cluster histogram, since the cluster 
has been shifted to higher magnitudes.  

Before performing the functional fit to our shifted data, we apply one of three observational biases 
to our $M_g$ histogram:  The cluster detection efficiency detailed in Section 4, the stellar detection 
efficiency described in \citet{nyetal02}, and 100\% detection efficiency (in which no correction is applied).
The first bias system reveals the evolution with increasing distance of globular cluster turnoff 
distributions as observed in SDSS data. The second bias system will produce the evolution of non-cluster 
turnoff distributions as a function of distance.  The final system reveals the evolution of turnoff 
distributions if no detection bias is applied, that is, if all of the stars originally detected in a 
cluster continue to be detected as the distances increase.

All ten clusters fit in Section 5 were distance shifted to increasingly greater $d_{\rm eff}$ values, 
using the methods outlined above.  At each $d_{\rm eff}$, a new set of fit parameters for the observed 
absolute magnitude distribution were evaluated using the methods described above.  An example of the 
histogram and fit of cluster M3 (NGC 5272), shifted to a $d_{\rm eff}$ of 44.0 kpc, is presented in the 
lower plot of Figure 6.

We present the results of the distance shifted fits for the parameters $\mu$, $\sigma_l$, and 
$\sigma_r$ in Figures 12, 13, and 14 respectively.  NGC 5053 and M15 (NGC 7078) are plotted with a red 
dotted series to indicate their status as expected outliers, while NGC 4147 is plotted with a blue 
dotted series to indicate that it contains few stars and therefore the fits contain large errors.  

\subsection{Observed vs. Intrinsic Absolute Magnitude Distribution of F Turnoff Stars}

It is important to note that all of the clusters studied, including suspected outliers, have similar 
$\mu$ and $\sigma_l$ values despite differences in distance, age and metallicity.  Although we see 
differences in $\sigma_r$, we have shown these are due to photometric errors, and not differences 
in the absolute magnitude distribution of turnoff stars in globular clusters.  This implies that 
the halo cluster population is intrinsically similar throughout.  In the next section we will show 
that this similarity is related to the Age-Metallicity Relationship.

In Figure 15 we show a series of fits to the turnoff star magnitude distribution for nearby cluster NGC 
6205 at increasing effective distances, including the effects of the cluster detection efficiency.  
From this plot, the most obvious effect is the loss of stars with distance; however, one can see that 
these losses balance to cause $\sigma_l$ to stay constant throughout, and $\mu$ shifts slightly to the 
left (brighter magnitudes) as the detection efficiency cuts in with distance.

We find that $\mu$ is approximately constant with distance, regardless of the applied detection 
efficiency bias.  From the plot of $\mu$ vs $d_{\rm eff}$ (Figure 12), we see that $\mu$ values decrease 
slowly with increasing distance; however, $\mu$ fit errors increase quickly with distance due to the loss 
of turnoff stars.  Within the fit errors, $\mu$ is adequately described by a constant value.  The 
error-weighted constant-value fit to $\mu$ gives $\mu=$ \mufit ($\pm$ \muerror), with a cluster dispersion 
of 0.073.  We also provide a linear fit to $\mu$ with distance, and find that a slope of zero (which would 
imply a constant value) is within the one-$\sigma$ errors:

\begin{equation}
	\mu(d_{\rm eff}) = -0.011(\pm 0.02) d_{\rm eff} + 4.39(\pm 0.57)
\label{mulinearfit}
\end{equation}

We find that $\sigma_l$ values also stay constant with distance, regardless of the applied detection 
efficiency, to within calculated errors (Figure 13).  It is also apparent that the two expected 
outlier clusters NGC 5053 and NGC 7078 have $\sigma_l$ behaviors that differ from the other clusters.
An error-weighted average to the $\sigma_l$ values, excluding the two outliers, give $\sigma_l=$ 
\siglfit ($\pm$ \siglerror), with a cluster dispersion of 0.18.  

The values of $\sigma_r$ do not stay constant with the distance shifts for any of our three detection 
efficiency cases.  Because of increasing color errors with distance, all but the nearest turnoff star 
populations are contaminated by redder main-sequence stars, as described in Section 6.2.  These redder 
stars enter the turnoff histograms on the fainter ($> \mu$) side, thereby widening the overall distribution 
and increasing $\sigma_r$ while leaving the other two parameters unchanged.  The nearest clusters (NGC 5272, 
NGC 5904, NGC 6205, NGC 7089; excluding the core-collapsed NGC 7078) are near enough that they do not 
exhibit significant red main-sequence contamination, and are consistent with each other.  We report that 
these nearby, uncontaminated clusters are representative of the ``true" globular cluster distribution, with 
a $\sigma_r$ fit of \gammafit ($\pm$ \gammaerror), equivalent to the fit value of $\gamma$, below.

When we apply the SDSS detection efficiency for globular clusters, we find that all of our clusters show 
the same behavior as a function of distance (Figure 14, left).  The initial, quick rise in $\sigma_r$ with 
distance is due to a large influx of red main-sequence stars due to color errors.  If the cluster is 
observed at even farther distances, $\sigma_r$ is reduced due to the cluster detection efficiency removing 
increasingly larger portions of the faint edge of the turnoff.  This consistent series in $\sigma_r$ is 
evidence that the observed spread in initial cluster $\sigma_r$ values is not a real feature of the clusters, 
but is instead an observational bias due to the incompleteness of SDSS crowded-field photometry at faint 
magnitudes.  In Figure 16, we show a fit to the variation of $\sigma_r$ with distance.

In order to see what happens to the distribution of of turnoff stars as a function of distance for
SDSS field stars, we also apply the SDSS stellar detection efficiency when distance shifting the cluster.  
We find that the nearby clusters (NGC 5272, NGC 5904, NGC 6205, NGC 7089, and NGC 5466) follow a similar 
pattern as in the cluster detection efficiency system, while the initially more distant clusters 
(NGC 4147, NGC 5024, NGC 5053, Pal 5) maintain a relatively constant value for $\sigma_r$.  The initially 
more distant clusters are already incomplete due to the cluster detection efficiency, which has modified 
their apparent $\sigma_r$ fit.  For the nearby clusters, the initial rapid rise in $\sigma_r$ with 
distance peaks at a higher value of $\sigma_r$, and the subsequent gradual decline is less severe.  
This is due to the wider and fainter drop off for the stellar detection efficiency as compared to 
the cluster detection efficiency; that is, the stellar detection efficiency starts removing fainter 
stars at greater distances, and to a less severe degree.  As the stellar field detection efficiency 
is the dominant observational bias in the SDSS, we provide a 4th order polynomial fit to $\sigma_r$
versus distance, with the coefficients given by $\mathbf{a_{\rm sdss}}$:

\begin{equation}
  \mathbf{a_{\rm sdss}} = (-1.7, 0.46, -0.02, 0.00057, -4.7\times 10^{-6})
\label{stellarfit}
\end{equation}

If we assume 100\% detection efficiency at all magnitudes, we find that nearby clusters see a quick rise 
in $\sigma_r$ as in the previous two cases, but then level out at a constant value (Figure 14, right; 
Figure 16), while farther clusters remain constant (as in the previous case).  As discussed in Section 6.2,
 the ratio of ``true" turnoff stars to red contaminants remains constant after the initial influx of red 
stars into the turnoff color cut.  Since this ratio remains constant, there is no appreciable change to 
the observed turnoff distribution after the initial rise.  Figure 16 shows the difference in $\sigma_r$
evolution for different detection efficiency cases.  

The sudden inflow of red turnoff contaminants between 10.0 and 15.0 kpc is responsible for the 
rapid rise in $\sigma_r$ fits for all three detection efficiency systems.  The nearby clusters (NGC 5272, NGC
5904, NGC 6205, and NGC 7089) are close enough ($\leq 11.5$ kpc) to not contain significant turnoff contamination; 
they are therefore representative of the intrinsic $\sigma_r$ value for globular clusters.  When we assume 
100\% detection efficiency at all magnitudes, the deviation of $\sigma_r$ fits from this intrinsic value is then
purely an effect of the color errors due to magnitude, and are not influenced by pre-existing red star 
contamination or detection efficiency losses.  These $\sigma_r$ trends then serve as a basis for understanding 
the effects of color errors on the observed distribution of turnoff stars, and so we provide a sigmoid functional 
fit to the 100\% detection efficiency case:

\begin{equation}
\sigma_r = \frac{\alpha}{1 + e^{- (d_{\rm eff} - \beta)} } + \gamma
\label{sigmoid}
\end{equation}

Where $d_{\rm eff}$ is the effective distance of the shifted cluster.  We find best fit values for the 
sigmoid functional fit to the nearby clusters are $\alpha=$ \alphafit ($\pm$ \alphaerror), $\beta=$
\betafit ($\pm$ \betaerror), and $\gamma=$ \gammafit ($\pm$ \gammaerror).  The three different trends 
for $\sigma_r$ with distance, produced by the three separate detection efficiency systems, are compared 
in Figure 16.

We have shown that the intrinsic turnoff distribution, as quantified by the fit parameters $\mu$, $\sigma_l$, 
and $\sigma_r$, is similar for all observed clusters, but that the measured value of $\sigma_r$ depends 
on the distance to the cluster due to photometric errors that increase if the cluster is farther away.
Stars that affect $\sigma_l$ are bright, and therefore have smaller color errors than stars that affect 
$\sigma_r$, and will therefore not be shifted into or out of the turnoff color cut as easily.  Also, 
there are few stars adjacent to the brighter region of the turnoff, so potential contamination is low.  
Finally, detection efficiency biases must effect the fainter turnoff stars before they can affect the 
brighter stars.  Therefore, bright turnoff stars are well insulated from contamination and detection bias.  
The dominant process affecting $\sigma_l$, then, is the loss of turnoff stars due to color errors.  Turnoff 
star losses also affect $\mu$, but at large distances $\mu$ will also be modified by detection biases.

\section{The Age-Metallicity Conspiracy}  

The age-metallicity relationship (AMR) describes the observed relationship between a cluster's age 
and its average metallicity.  Recent observational work \citep{DA2005, MF2009, D2011} indicates that 
there are two such AMRs present in the Milky Way: one in which metallicity increases with age, 
attributed to clusters and dwarf galaxies that were that were gradually captured by the Milky 
Way over time (considered by these authors to be ``outer halo" clusters); and a set of clusters with 
age ~13 Gyr that spans a very large range of metallicities, believed to be due to old clusters that 
formed rapidly alongside the Milky Way during the initial formation event (considered to be ``inner 
halo" clusters).  

In Figure 17, we show that the clusters in this study follow the trend of metallicity decreasing with 
age, which we will henceforth refer to as \textit{the} AMR, as presented in \citet{mg10}.  Muratov \& Gnedin 
provide a galaxy-independent model of galaxy formation history using semi-analytic models that take into 
account cosmological simulations.  Note that the only two clusters in our sample that are not good matches
to Muratov \& Gnedin have already been identified as outliers in previous discussions.  Since all of our 
clusters are located at high galactic altitudes, and are likely members of the galactic halo (and though 
to have been accreted during galaxy assembly), it is not surprising that our clusters are similar to 
the \citet{mg10} relations.  Figure 17 shows that the Muratov \& Gnedin AMR model does not reproduce the 
constant (old, rapid co-forming clusters) metallicity trend described by \citet{DA2005}, \citet{MF2009}, 
\citet{D2011}\footnote{\citet{D2010} used Zinn \& West metallicities; these were converted to the 
Carretta \& Gratton scale in Figure 17 to be consistent.}, but that it is consistent with the AMR (young, 
late accretion) cluster values from this work and other papers.

Using the Muratov \& Gnedin AMR and the modified Padova isochrones, we show that they predict the 
absolute magnitudes of turnoff stars will be similar for old stellar populations that follow the 
AMR.  Taking the approximate mean metallicity at range of ages over 8 Gyr from Figure 8 of \citet{mg10}
(plotted in Figure 17 as the solid line), we produce a set of metallicity versus age points that are 
representative of the AMR.  We show the Padova isochrones (modified by the linear correction 
function in Equation 1) in Figure 18.  The blue turnoff point can be found between 
$0.20 < g-r < 0.23$, the turnoff magnitude lies between $3.77 < M_g < 4.16$, and the subgiant branch 
is constrained between magnitudes of $3.4 < M_g < 3.6$.  Also shown in Figure 18 are two isochrones 
generated for age-metallicity combinations far from the AMR.  These illustrate that the tight grouping 
of the series is a property of clusters on AMR, and is not true for arbitrary combinations of age and 
metallicity.

The turnoff parameter fits to the unmodified cluster data in Table 2 are consistent with the trends 
seen in Figure 18.  Along the AMR, as age decreases and metallicity increases, we see the turnoff move 
to slightly brighter magnitudes (lower $\mu$) and slightly redder colors (higher $(g-r)_0$).  We find 
that the AMR produces clusters with similar isochrones, but the age of a cluster slightly dominates 
over the metallicity in determining the $g_0$ turnoff brightness, while the metallicity slightly dominates 
over the age in determining the $(g-r)_0$ turnoff color.  

This finding implies that age and metallicity values along the AMR ``conspire" to produce a similar 
population distribution for all older ($8.0+$ Gyr) clusters.  As a cluster ages, it's turnoff becomes 
redder since it has had more time for stars to evolve off of the main-sequence.  However, an increase 
in metal content also corresponds to a redder value for the entire isochrone.  The plot of isochrones 
along the theoretical AMR in Figure 18 implies that these two effects almost exactly cancel each other 
out, and therefore old stars formed in accordance with the AMR should fall in a very narrow range on a 
color-magnitude diagram.  Provided one does not select a cluster from the old (``inner halo") population,
this finding simplifies distance measurements for the predominantly old stellar population in the 
Galactic halo, but complicates the age determination process, which depends on the uniqueness of 
isochrones.  In this paper, we have removed this complication by using only metallicities determined 
from spectra; only age and distance were used free parameters in our isochrone fits.

\section{Application to SDSS Data}

In this paper we have described the SDSS photometric errors and the resulting effects on observed F turnoff 
distributions ($0.1 < (g-r)_0 < 0.3$). In this section we will provide a method by which SDSS
observations can be corrected for these effects.  If we sum the polynomials whose coefficients are
presented in Equation 9 and Equation 10, we will produce a function that describes the ratio of stars 
selected as turnoff stars to the number of actual turnoff stars present, as a function of distance
\begin{equation}
  \varepsilon(r) = \frac{n(r)}{n_0} = \sum_{i=0}^{7} (\mathbf{a_{yi}}(r) + \mathbf{a_{ri}}(r))r^i
\label{densfix}
\end{equation}
where $r$ is the distance to the stellar population. Note that this equation assumes a $(u-g)_0$ cut to 
the data; if this cut is not performed, then additional stars will be included in the data set and this 
equation will not be valid.  Instead, the following equation should be used:
\begin{equation}
  \varepsilon(r) = \frac{n(r)}{n_0} = \sum_{i=0}^{7} (\mathbf{b_{yi}} + \mathbf{b_{ri}})r^i
\label{nocutfix}
\end{equation}
\begin{equation}
  \mathbf{b_{yi}} = (1.02, -0.011, -0.00043, 9.7\times 10^{-6},  -5.5\times 10^{-8}) \\
\label{nocuty}
\end{equation}
\begin{equation}
  \mathbf{b_{ri}} =( 0.034,  -0.039,   0.011, -0.00071, 2.0 \times 10^{-5},  -2.9 \times 10^{-7}, 
		2.1\times 10^{-9}, -6.4\times 10^{-12})
\label{nocutr}
\end{equation}

For simple density searches that ignore F turnoff distribution statistics, dividing the observed 
density at a given distance by $\varepsilon(r)$ is sufficient to correct data for missing turnoff
stars due to magnitude errors.  For example, if one measures the number of background-corrected F turnoff 
stars in a section of the Sagittarius tidal debris stream at a distance of 45 kpc, that number should be 
corrected for completeness (if necessary) and then divided by $\varepsilon(45$ kpc$)$ in order to produce 
the true number of turnoff stars in that field. 

For statistical models that seek to quantify F turnoff star densities in the Galactic halo, a more 
complicated correction method is required.  In \citet{cole08}, the authors sought to map the Sagittarius
tidal debris stream by performing a maximum likelihood fit of a statistical model to F turnoff stars 
taken from the SDSS data.  They include the effect of the turnoff star distribution by convolving a 
normalized Gaussian, representing the turnoff star distribution (mean=4.2, $\sigma=$0.6), with a 
halo density model in which all of the turnoff stars have the same absolute magnitude ($M_g=$4.2).  
A correction for the stellar detection efficiency is subsequently applied.

We seek to modify the methods of \citet{cole08} by incorporating our results into their analysis. In 
order to include the turnoff star distribution presented in this paper, the turnoff distribution 
in the density convolution must be changed from a symmetrical Gaussian ($\mathcal{N}$ in that paper) 
to the double-sided Gaussian, $\mathcal{G}(g_0; \mu, \sigma_l, \sigma_r, A)$ presented in Equations 4 
and 5, with a $\sigma_r$ given by Equation 13.  Note that we use the 100\% detection efficiency 
function for $\sigma_r$, as the stellar detection efficiency is applied subsequent to the turnoff 
convolution. The normalization for $\mathcal{G}$ must be adjusted for the fact that turnoff stars are 
lost (and a smaller number of G stars are gained) at farther distances.  To account for the turnoff 
dissipation and contamination, the distribution $\mathcal{G}$ should be multiplied 
by $\varepsilon(g)$.  Putting this together, we produce a convolution recipe which provides 
a statistical description of the Galactic halo turnoff population, including the effects of the 
double-sided Gaussian magnitude distribution and the effects of dissipation and contamination 
with distance.  Equation 14 in \citet{cole08} is replaced by:
\begin{equation}
\rho(g_0, \Omega) = 
\int_{-\infty }^{\infty} dg \mbox{ } 
\varepsilon\left(  r(g, \mu) \right)  
\rho_{\mu}(g, \Omega) 
\mathcal{G} \left(  g_0-g; \mu, \sigma_l, \sigma_r(r(g,\mu))\right)
\label{distfix}
\end{equation}
where $\rho$ represents a density function of magnitude and solid angle ($\Omega$),  
$\rho_{\mu}(g, \Omega)$ is the model turnoff density function, in which all stars are assumed to 
be at absolute magnitude $\mu$ (\mufit), $\varepsilon$ is Equation 14, and $\mathcal{G}$ is Equation 4 
with Equation 13 as the input for $\sigma_r$.  $\varepsilon$ and $\sigma_r$ are both functions of
$r$, which is equal to the distance ($r=10^{\frac{g_0 - \mu - 10}{5}}$).

Using these examples as templates, and with the information contained in the rest of this paper,
future authors should be able to adapt our results to correct their models for observational biases
due to turnoff magnitude errors.  Future authors must be prepared to compensate for three mechanisms 
that cause turnoff star incompleteness and misidentification:  The combined effects of dissipation 
and contamination, given by Equation 14; the SDSS stellar field detection efficiency, given in 
\citet{nyetal02}; and the intrinsic magnitude distribution of turnoff stars within the Galactic halo,
as discussed in Section 6 and this section.

\section{Discussion}

\subsection{Effects of Abundance Variation}

Recent observational studies \citep{pritzl2005, r2009} (see also reviews by Gratton et al. 2004 and
McWilliam 1997) indicate that both outer halo globular clusters and halo field stars have similar 
alpha-element abundances, consistent with an average [$\alpha$/Fe] value of ~0.3.  This is strong evidence
that halo clusters and field stars share a similar formation history, and it is therefore reasonable to
assume that they share similar population distributions and photometric properties.  These studies also 
find that halo clusters and stars are chemically distinct from dwarf spheroidal galaxies \citep{venn2004} 
and the galactic disk, including stars in the thick disk.  While thick disk stars can be removed from SDSS 
photometric data through magnitude and color cuts, dwarf spheroidal stars may be present in the halo 
due to tidal disruptions of infalling dwarfs.  If significantly different [$\alpha$/Fe] affect the photometric
properties of a population, disrupted dwarf spheroidal galaxies may pose a problem for our technique.
However, the success of \citet{cole08} in mapping segments the Sagittarius tidal stream with a similar
technique is evidence that that dwarf galaxy turnoff star distributions are not significantly different from
those in the globular clusters in this study.

The set of globular clusters in this study are limited to old, metal poor galactic halo clusters, with 
a maximum metallicity of [Fe/H]$=-1.17$ and a minimum age of 9.5 Gyr (NGC 5904).  Figure 18 indicates 
that our results can be extended to an age of at least 8.0 Gyr.  From Figure 17 we see that our entire 
cluster sample falls within the ``metal poor" (blue) region of the Muratov \& Gnedin AMR.  While our 
results are well-suited to our goal of describing the old, metal-poor halo of the Milky Way, we have not 
tested how far it can be extrapolated to stellar populations less than 8 Gyr old or more metal-rich than 
NGC 5904.  

The results of this paper may not apply to globular clusters with age and metallicity values that fall 
far from the Milky Way AMR.  Our results may not apply to stellar populations in other galaxies, which 
have differing assembly histories and potentially different age-metallicity relationships.  Indeed, the
two unusual globular clusters that fall outside of the AMR in Figure 17, are the two clusters that differ 
from each other in turnoff magnitude (Figure 11).

Recently, multiple star formation periods have been detected in globular clusters \citep{mil2008,
bedin04, piotto2007} (see review by \citet{piotto2009}). We do not expect multiple stellar populations to 
significantly affect our results, however, certain clusters with several separate, strongly visible main 
sequences (such as M54) may be poorly described by our results.

Above average helium enrichment has been proposed as a potential mechanism for producing separations in 
multiple-population cluster main sequences \citep{piotto2007, mil2008, pas2011}.  Due to the difficulty of 
measuring He enrichment, the complete effects of enrichment are still under investigation (see preliminary 
work by Valcarce at al. 2011, and Milone \& Merlo 2008), but it is thought that it will result in slightly 
brighter subgiant branches and slightly bluer turnoffs, since He is less opaque than H.  If the He abundances
have only a small effects on photometry, they will not significantly influence our results.

\subsection{Potential Influence of Binaries}

Binary stars present in globular clusters have the potential to bias color-magnitude diagrams towards
brighter, redder values, as detailed in \citep{rw1991}. The maximum increase in brightness occurs when
the binaries are of the same spectroscopic type, $2.5*log(2)\approx 0.75$ magnitude, and the maximum 
increase in ``reddening," which serves to widen the color-magnitude diagram towards redder values, 
can shift colors by as much as 0.05 magnitudes.  Depending on the binary fraction ($f$), this effect 
could introduce significant biases when determining population statistics of a globular cluster.

The binary fraction in old globular clusters is generally low, around a few percent to 20\%, with the 
binaries being concentrated near the cluster's center \citep{sol2007, fetal09}, although a 
select few globular clusters have $f$ values as high as 50\%.  The primary result of Monte-Carlo simulations 
in \citet{fetal09} is that ``True" binary fractions are likely to remain constant with age, implying that 
we can also rule out age-dependent biases on cluster $f$ values.  \citet{carn2003} studied the binary 
fractions of metal-poor ([Fe/H] $\leq$ -1.4) field red giants and dwarfs, and found them to be 
16\%$\pm$4\% and 17\%$\pm$2\%, respectively, which is on the high end of a typical globular cluster $f$.
Globular clusters also feature apparent binaries, which are stars that appear as binaries due to 
crowding, which become more likely as one looks closer to the densely-packed cluster cores.

Due to the poor performance of the SDSS photometric pipeline in crowded fields, the centers of globular 
clusters are absent from the database, and therefore are not present in this study.  If binaries are 
concentrated towards a cluster's center as in \citet{sol2007, fetal09}, then this ``disadvantage"
serves to reduce the potential bias that binaries would have on our results, and will also greatly reduce 
the number of apparent binaries in our data.  

The effect of a binary population is to shift the observed turnoff to brighter magnitudes, and to increase 
the spread in magnitudes.  We ran an experiment in which we generated 1000 stars with magnitudes given by a 
double-sided Gaussian with $\mu=4.2$, $\sigma_l = 0.36$ and $\sigma_r = 0.76$, and with colors given by a 
standard Gaussian with a $\mu_{(g-r)}=0.25$ and $\sigma_{(g-r)} = 0.025$.  We then randomly selected 20\% of 
these stars and shifted them brighter by 0.75 magnitudes and redder by 0.05 magnitudes, which should produce a 
larger effect than we would expect from the binary fraction of a typical halo globular cluster.  We found that 
the 20\% binary population had a new magnitude fit of $\mu=4.22$, $\sigma_l = 0.55$ and $\sigma_r = 0.75$, and 
a color fit of $\mu_{(g-r)}=0.26$ and $\sigma_{(g-r)} = 0.03$.  The variations in these values are all within 
the errors of the fitting process, except for $\sigma_l$ in the magnitude fit.  It is possible that the quantity 
$\sigma_l$ is sensitive to to the binary fractions typically found in halo globular clusters, in the sense that
a large binary fraction could produce a larger $\sigma_l$.  The spread in $\sigma_l$ values in our accepted 
sample is low (Figure 13), so we do not expect that the effect of binaries has changed our results significantly. 
A future study could potentially use a more rigorous study of binary fraction 
effects to see if the binary fraction is directly correlated to the value of $\sigma_l$.

We expect the results of this paper to provide a close approximation of the F turnoff absolute magnitude 
distribution in old, metal poor stars in the Milky Way halo, as observed by SDSS.  These results should 
also be a reasonable approximation of small dwarf galaxy turnoff distributions.  We do not expect the 
results to be applicable to young, relatively metal-rich stellar populations, to populations that are outliers
on the AMR, or to populations that are outside of the Milky Way.

\section{Conclusions}

In this paper we analyze SDSS data for 11 globular clusters in the halo of the Milky Way, and draw 
conclusions about the intrinsic and observed absolute magnitude distributions of F turnoff stars.  The
major conclusions are:

(1) The completeness as a function of magnitude in SDSS stellar data is different in crowded fields 
such as globular clusters than in typical star fields.  This incompleteness begins at brighter 
magnitudes and falls off more steeply than the standard stellar detection efficiency reported in 
\citet{nyetal02}.  We calculated an approximate parabolic function to describe this crowded-field 
incompleteness, as presented in Equation 3.

(2) We calculated a linear correction to the turnoff region of Padova isochrones in SDSS colors in order 
to place them on an age scale that is consistent with determinations from other authors.  The findings of 
\citet{an2009} were used as a basis for this function.  Without this correction, Padova \textit{ugriz} 
isochrones imply ages greater than the current measured age of the universe.  We find the appropriate 
correction to be $\Delta(g-r) = -0.015*M_g + 0.089$, valid for $3.5 < M_g < 8.0$.

(3) Using our Padova isochrone correction, a uniformly determined set of metallicity ($[Fe/H]$), distance 
(kpc), and age (Gyr) measurements were calculated for 11 globular clusters observed in the Sloan
Digital Sky Survey.  These results are collected in Table 1.

(4) As color errors increase with distance, the fraction of F turnoff stars detected in a narrow color
range decreases rapidly, while redder star contamination becomes significant.  Using Equations 9 
and 10, future authors will be able compensate for both effects.

(5) Across a range of old stellar populations ($-2.30 < [Fe/H] < -1.17$, dex; $13.5 <$ Age $< 9.5$, Gyr), 
the distribution of SDSS $g$ filter absolute magnitudes of the turnoff stars ($0.1 < (g-r)_0 < 0.3$) is 
approximately constant:  $\mu = M_g=$ \mufit $\pm$ \muerror (dispersion=0.073), and if the distribution 
is fit with two half Gaussians the brighter half Gaussian has $\sigma_l =$ \siglfit $\pm$ \siglerror 
(dispersion=0.18), and the fainter half Gaussian, when not affected by observational errors, has 
$\sigma_r =$ \gammafit $\pm$ \gammaerror.  

(6) Although the $M_g$ turnoff absolute magnitude distribution is intrinsically the same for old stellar 
populations in the Milky Way, observational errors produce biases which introduce a significant difference 
in the absolute magnitude distribution for stars selected in a narrow color range.  While the average 
and bright-side half Gaussian fit parameters stay relatively constant to within errors, 
the observed fainter-side half Gaussian width ($\sigma_r$) changes dramatically with increasing 
distance (and therefore increasing errors) and depends strongly upon observational biases.  We 
found that Equation 13 provides a good description of the change in $\sigma_r$ with distance, as 
influenced by color errors only.  This must be accounted for if F turnoff color cuts are to be used 
to trace stellar structure in the Milky Way.

(7) The turnoff properties for old stellar populations is consistent with the known Age-Metallicity 
Relationship for the Milky Way.  Older globular clusters should have fainter, redder turnoffs, but 
because they are more metal-poor (which tends to produce brighter, bluer turnoffs) the turnoff is 
similar for all older clusters.  For stellar populations at least 8.0 Gyr old, the turnoff color 
ranges from $0.20 < (g-r)_0 < 0.23$, the turnoff magnitudes range between $3.77 < M_g < 4.16$, 
and the subgiant branch is constrained between the magnitudes of $3.4 < M_g < 3.6$.

(8)  We provide a method for correcting SDSS data for the effects of magnitude errors.  We discuss 
two cases:  (i) If the stellar population under study is at a single distance, then the observed F 
turnoff star number counts should be corrected for completeness, and then divided by $\varepsilon(r)$,
as given by Equation 14 or 15;  (ii)  If the stellar population is distributed in distance, one can 
convolve the proposed density model with a function that accounts for observational errors before
comparing with data.  

F turnoff stars have proven useful as tracers of Galactic structures, particularly in the halo.  
Before completing this work, we were concerned that the assumption of a single absolute magnitude
distribution for turnoff stars in the halo would be problematic.  Instead, our analysis shows that 
while a single turnoff is a good assumption, photometric errors significantly affect the stellar 
population as a function of apparent magnitude.  Since we have found that all old stellar 
populations are intrinsically similar, it is possible to model the selected populations of stars 
(as selected in a narrow color range) as a function of magnitude and correct measurements of 
Galactic structure for this effect.

\acknowledgments

Funding for this research was provided by the National Science Foundation, under grant AST 10-09670. 
Matthew Newby was partially supported by the NASA/NY Space Grant.  Heidi Newberg was partially 
supported by NSFC grants 10973015 and 11061120454.  Jacob Simones was supported by an NSF REU 
supplement to IIS 06-12213.  Matthew Monaco was supported by NSF REU grant DMR 08-50934.  

We would like to thank our anonymous referee for providing helpful feedback that improved the results 
of this paper, and made our conclusions more robust.  We would also like to acknowledge Brian Yanny for 
providing guidance with navigating the SDSS database, and to Benjamin Willett, David Horne and Jeffery 
Carlin for providing advice and discussion that helped move this project forward.

Funding for the SDSS and SDSS-II has been provided by the Alfred P. Sloan Foundation, the 
Participating Institutions, the National Science Foundation, the U.S. Department of Energy, the 
National Aeronautics and Space Administration, the Japanese Monbukagakusho, the Max Planck Society, 
and the Higher Education Funding Council for England. The SDSS Web Site is http://www.sdss.org/.

The SDSS is managed by the Astrophysical Research Consortium for the Participating Institutions. 
The Participating Institutions are the American Museum of Natural History, Astrophysical Institute 
Potsdam, University of Basel, University of Cambridge, Case Western Reserve University, University 
of Chicago, Drexel University, Fermilab, the Institute for Advanced Study, the Japan Participation 
Group, Johns Hopkins University, the Joint Institute for Nuclear Astrophysics, the Kavli Institute 
for Particle Astrophysics and Cosmology, the Korean Scientist Group, the Chinese Academy of Sciences 
(LAMOST), Los Alamos National Laboratory, the Max-Planck-Institute for Astronomy (MPIA), the 
Max-Planck-Institute for Astrophysics (MPA), New Mexico State University, Ohio State University, 
University of Pittsburgh, University of Portsmouth, Princeton University, the United States Naval 
Observatory, and the University of Washington.

\clearpage

\begin{deluxetable}{lccccccccr}
\tabletypesize{\scriptsize}
%\rotate
\tablewidth{0pt}
\tablecolumns{10}
\tablecaption{ Results of modified Padova isochrone fits}
\tablehead{
\colhead{NGC}    & \colhead{Messier} & \colhead{$l$}          & \colhead{$b$}          & \colhead{Metallicity} & 
\colhead{Fit Distance} & \colhead{Fit Age $\pm$ Error}  & \colhead{$r_{clus}$} & \colhead{$r_{cut}$} & \colhead{\# Stars in} \\
\colhead{Number} & \colhead{Number}  & \colhead{${}^{\circ}$} & \colhead{${}^{\circ}$} & \colhead{CG97 [Fe/H]}     & 
\colhead{(kpc)}   & \colhead{(Gyr)}   & \colhead{${}^{\circ}$} & \colhead{${}^{\circ}$}&  \colhead{Cluster\tablenotemark{1}} }
\startdata
NGC 4147 & \nodata & 252.85 & 77.19 & -1.58 & 19.3 & 11.2 $\pm$ 0.8 & 0.07 & 0.08 &    583\\
NGC 5024 & M53     & 332.96 & 79.76 & -1.89 & 18.7 & 12.5 $\pm$ 1.3 & 0.25 & 0.30 &   5871\\
NGC 5053 & \nodata & 335.70 & 78.95 & -2.30\tablenotemark{2} & 18.5 & 11.5 $\pm$ 1.1 & 0.14 & 0.17 & 2280\\
NGC 5272 & M3      & 42.22  & 78.71 & -1.43 & 10.4 & 12.2 $\pm$ 1.1 & 0.30 & 0.40 &   4395\\
NGC 5466 & \nodata & 42.15  & 73.59 & -2.14 & 15.6 & 13.4 $\pm$ 0.9 & 0.18 & 0.25 &   2029\\
NGC 5904 & M5      & 3.86   & 46.80 & -1.17 & 8.0  & 9.4 $\pm$ 0.1 & 0.21 & 0.28 &   3488\\
NGC 6205 & M13     & 59.01  & 40.91 & -1.42 & 7.7  & 12.7 $\pm$ 0.2 & 0.24 & 0.33 &   2192\\
NGC 6341 & M92     & 68.34  & 34.86 & -2.17 & 8.7  & 13.4 $\pm$ 0.9 & 0.135 & 0.20 &   334\\
NGC 7078 & M15     & 65.01  & -27.31& -2.04 & 11.0 & 11.2 $\pm$ 0.9 & 0.19 & 0.25 &   2939\\
NGC 7089 & M2      & 53.37  & -35.77& -1.38 & 11.5 & 12.6 $\pm$ 0.3 & 0.14 & 0.18 &   1451\\
Pal 5    & \nodata & 0.85   & 45.86 & -1.24 & 21.0 & 11.5 $\pm$ 0.7 & 0.087 & 0.12 &  1478\\
\enddata
\tablenotetext{1}{The number of stars represents the number of stars contained within $r_{clus}$, and therefore the number 
of stars used to build the cluster dataset.}
\tablenotetext{2}{The Zinn \& West metallicity of NGC 5053 (-2.58) was outside the effective conversion range of the CG97 
metallicity scale, as well as the range of available Padova isochrones, so the \citet{cbg09} value of -2.30 was used instead}
\end{deluxetable}

\begin{deluxetable}{lcccccccccccc}
\tabletypesize{\scriptsize}
\rotate
\tablewidth{550pt}
\tablecolumns{13}
\tablecaption{ Fit values and errors to cluster data \label{ gctab }}
\tablehead{
\colhead{NGC ID} & \colhead{Messier ID} & \colhead{FWHM} & \colhead{$\mu$} & \colhead{$\mu$ error} & \colhead{$\sigma_l$} 
& \colhead{$\sigma_l$ error} & \colhead{$\sigma_r$} & \colhead{$\sigma_r$ error} & \colhead{$A$} & \colhead{$A$ error} 
& \colhead{$(g-r)_{TO}$} & \colhead{$(g-r)_{TO}$ error} }
\startdata
NGC 4147 & \nodata & 1.70 & 4.42 & 0.33 & 0.50 & 0.22 & 0.98 & 0.29 & 26.6  & 3.34 & 0.223 & 0.022 \\
NGC 5024 & M53     & 1.69 & 4.49 & 0.02 & 0.50 & 0.02 & 0.94 & 0.03 & 271.8 & 3.53 & 0.240 & 0.010 \\
NGC 5053 & \nodata & 1.92 & 4.68 & 0.07 & 0.70 & 0.06 & 0.93 & 0.07 & 105.3 & 3.00 & 0.219 & 0.011 \\
NGC 5272 & M3      & 1.37 & 4.33 & 0.05 & 0.39 & 0.04 & 0.77 & 0.04 & 154.9 & 3.71 & 0.251 & 0.009 \\
NGC 5466 & \nodata & 1.92 & 4.28 & 0.15 & 0.40 & 0.09 & 1.22 & 0.13 & 78.7  & 3.17 & 0.217 & 0.010 \\
NGC 5904 & M5      & 1.23 & 4.13 & 0.28 & 0.32 & 0.18 & 0.72 & 0.19 & 134.7 & 4.03 & 0.253 & 0.006 \\
NGC 6205 & M13     & 1.30 & 4.31 & 0.13 & 0.31 & 0.09 & 0.79 & 0.10 & 69.0  & 4.07 & 0.251 & 0.008 \\
NGC 7078 & M15     & 2.18 & 4.48 & 0.11 & 0.62 & 0.08 & 1.23 & 0.11 & 88.7  & 3.26 & 0.219 & 0.011 \\
NGC 7089 & M2      & 1.27 & 4.33 & 0.10 & 0.31 & 0.09 & 0.76 & 0.12 & 46.0  & 4.07 & 0.258 & 0.010 \\
Pal 5    & \nodata & 1.56 & 4.21 & 0.14 & 0.33 & 0.10 & 0.99 & 0.12 & 54.1  & 3.55 & 0.284 & 0.011 \\
\tableline
Average  & \nodata   & 1.613 & 4.367 & 0.138 & 0.440 & 0.097 & 0.930 & 0.122 & 103.0 & 3.573 & 0.242 & 0.011\\
Standard & Deviation & 0.309 & 0.149 & 0.092 & 0.129 & 0.059 & 0.173 & 0.073 & 67.5  & 0.367 & 0.021 & 0.003\\
\enddata
\end{deluxetable}

\clearpage
\setcounter{page}{1}

\begin{figure}
\label{skyplot}
\figurenum{1}
\plotone{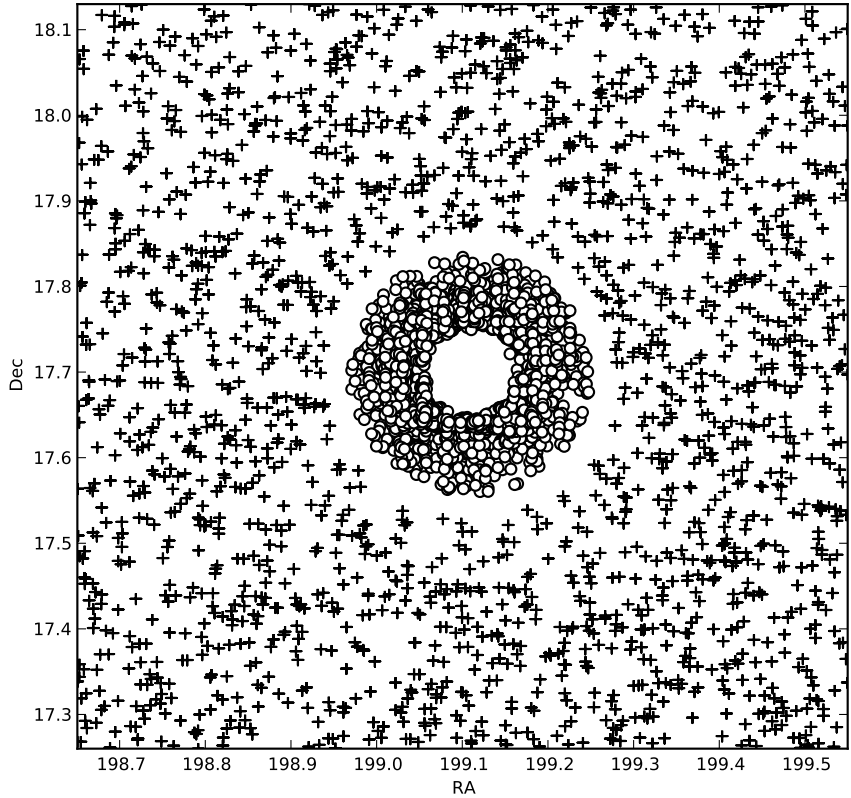}
%\epsscale{}
\caption{Positions of stars near globular cluster NGC 5053 in right ascension and declination. Stars 
assigned to the cluster are marked with circles, while stars assigned to the background are marked 
with crosses.  At the edge of the cluster, it is difficult to separate cluster stars from background 
stars, so a ring of stars at the interface has been removed, leaving a gap on the plot.  The empty 
area in the center of the plot is where the SDSS photometric pipeline was unable to separate individual 
stars, and the data was not included in the database.  This area has been subtracted from the cluster 
area determinations in Section 2.}
\end{figure}

\begin{figure}
\label{isofit}
\figurenum{2}
\plotone{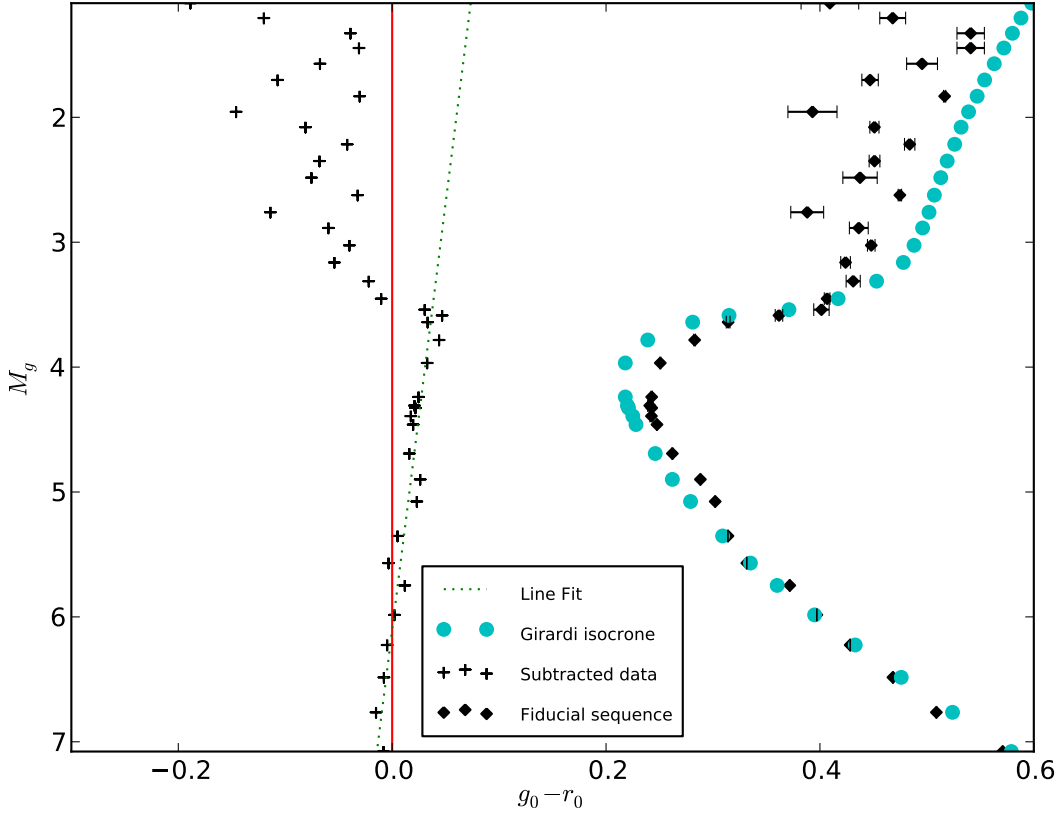}
%\epsscale{}
\caption{Diagram of Padova Isochrone fit for cluster NGC 6205.  The values of the raw Padova 
Isochrone, (blue circles) generated from the \citet{an2009} results, were subtracted from the values 
of the cluster fiducial fit (black diamonds) to produce the offset (crosses).  A red vertical 
line indicates zero offset, or a perfect match between isochrone and fiducial fit.  A linear 
fit (dotted line) to the offset was made for values below the giant branch ($M_g > 3.5$):  
$\Delta(g-r)_0 = -0.015*M_g + 0.089$.  This correction causes Padova isochrones to fit the turnoff on 
color-magnitude diagrams.  We then used this fit to match isochrones to our cluster sample (see text).}
\end{figure}

\begin{figure}
\label{isogroup}
\figurenum{3}
\plotone{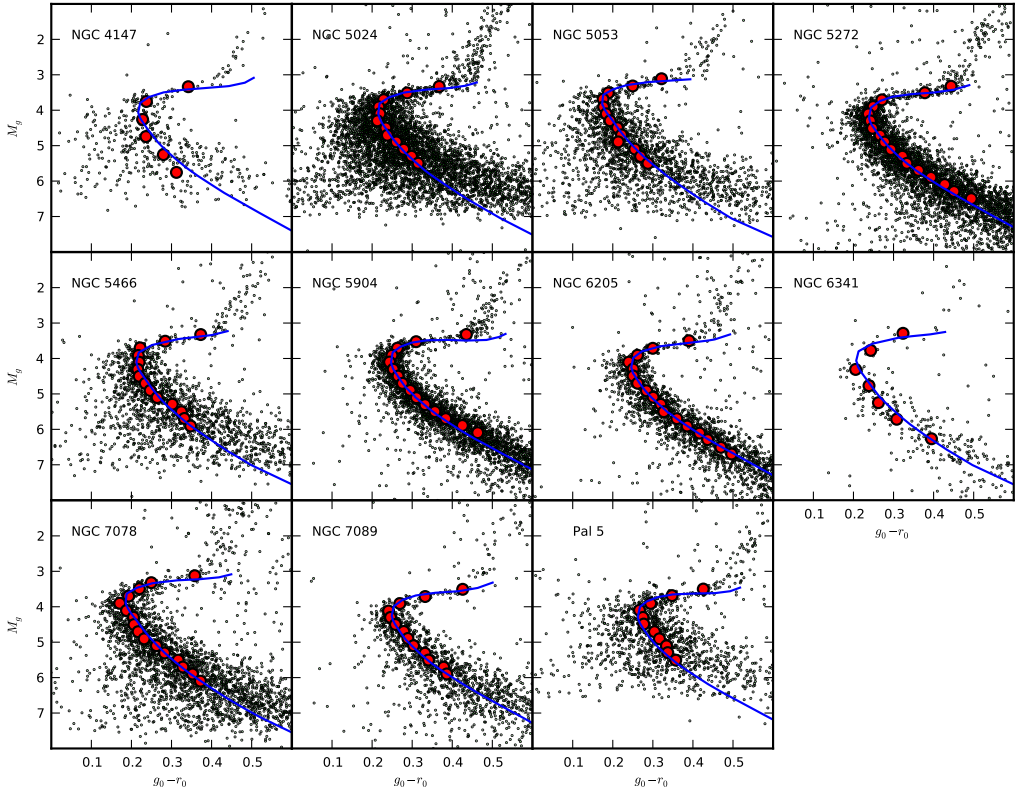}
%\epsscale{}
\caption{Color-magnitude diagrams, fiducial fits, and isochrone fits for all eleven globular 
clusters used in this study.  Black dots represent the positions of individual stars in 
$M_g$, $(g-r)_0$; red circles represent 2-$\sigma$ rejection fiducial fits to the stars in
$M_g$ strips; and blue lines represent Padova isochrones modified by our correction function.  
The ages, metallicities, and distances for the isochrones are given in Table 1.  Note the 
absence of faint stars in some clusters. This is due to the poor performance of the SDSS 
photometric pipeline in crowded fields.  The shape of this incompleteness is described in 
the text.  Also note the sparse data in NGC 6341; it will not be used for F turnoff analysis.}
\end{figure}

\begin{figure}
\label{delta_plots}
\figurenum{4}
\plotone{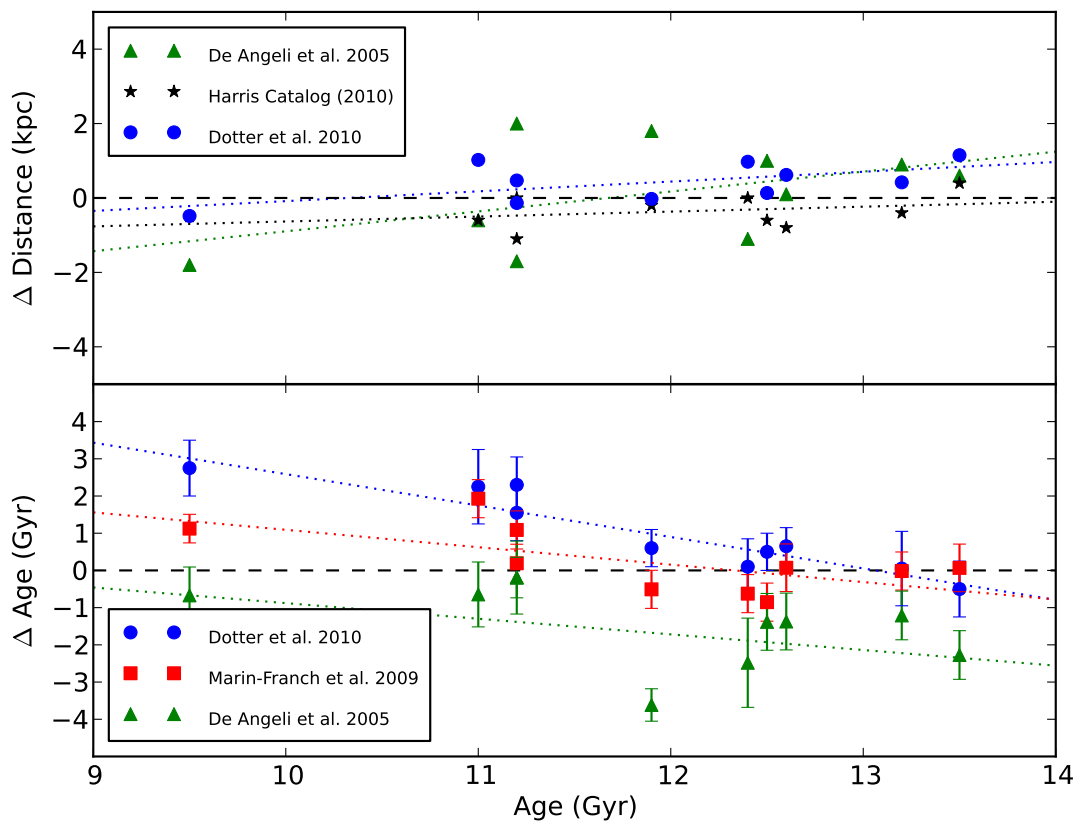}
%\epsscale{}
\caption{Comparison of the distance and age values from this paper with values from other sources.
The top panel plots the distances from this paper, subtracted from the respective external values, 
plotted vs. our fit ages.  Dotted trend lines are linear fits to the points.  Our distances are in 
good agreement with the previous literature.  The lower panel is the same as the top panel, but with 
the age differences as the y-axis.  Our ages agree with the previous literature to within the 
formal errors, but have a small linear offset.  Note that Palomar 5 is absent from this plot, as 
the other authors did not study this cluster.}

\end{figure}

\begin{figure}
\label{clusroll}
\figurenum{5}
\plotone{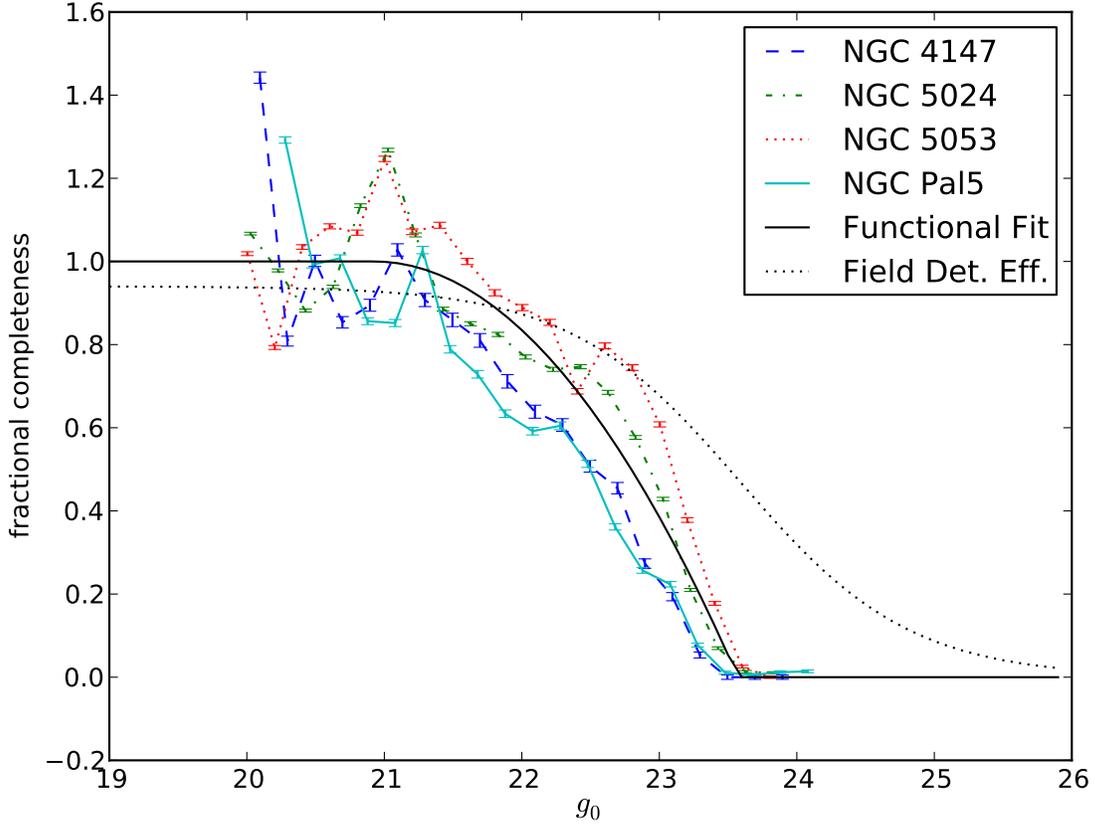}
%\epsscale{}
\caption{The SDSS crowded-field photometry detection efficiency.  Due to the poor performance 
of the SDSS photometric pipeline in crowded fields, more distant clusters have a deficit of 
stars at fainter magnitudes relative to brighter clusters.  The ratio between the number of 
stars observed in distant clusters and the reference histogram are plotted as colored lines.  
The reference histogram was created from the average histogram of $M_g$ for the nearby clusters 
NGC 6205, NGC 5904, and NGC 5272, and then shifted to the distance of each of the distant 
clusters NGC 4147, NGC 5024, NGC 5053, and Pal 5.  The parabolic functional fit 
($CDE(g_0)=1.0 - 0.14 \cdot (g_0 - 20.92)^2$, for $20.92 < g_0 < 23.57$) is plotted as the 
solid black line. The stellar detection efficiency from \citet{nyetal02} is plotted as a 
dotted series for reference.}
\end{figure}  

\begin{figure}
\label{histfit}
\figurenum{6}
\plotone{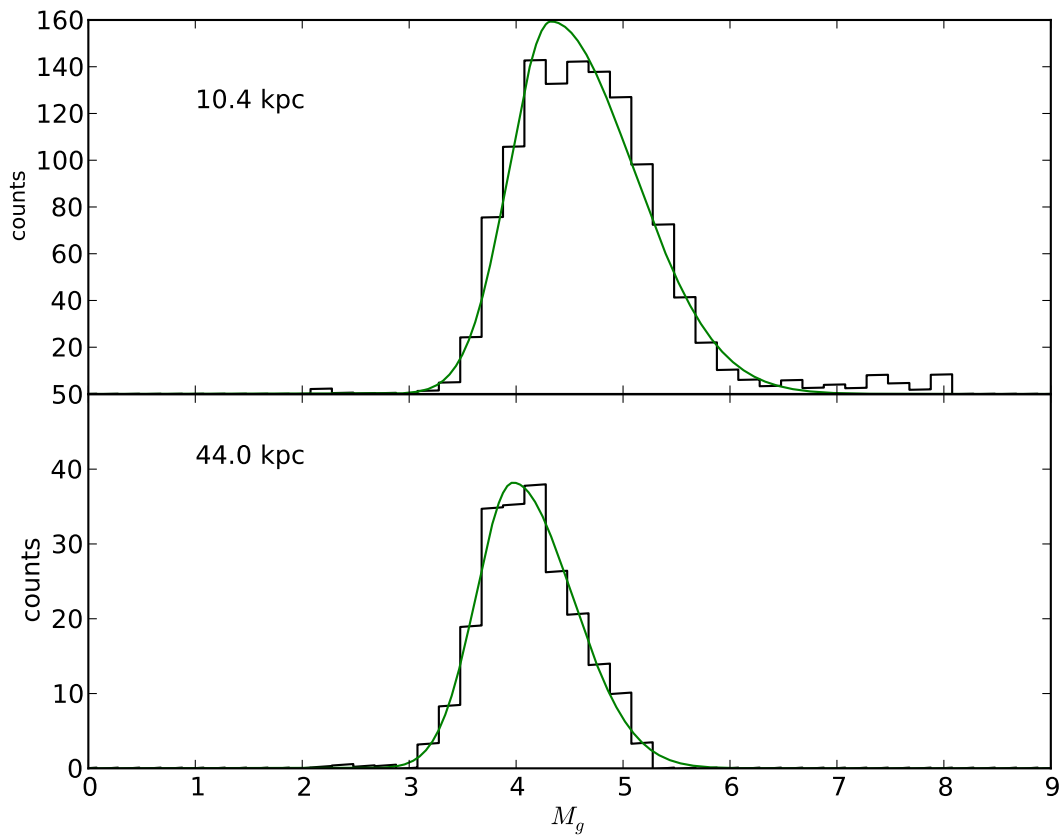}
%\epsscale{}
\caption{Histogram fits for NGC 5272 original data (top) and convolved to to an effective distance
of 44.0 kpc (bottom) with the cluster detection efficiency applied.  Blue lines outline the bins, while 
blue points mark the bin centers, to which a double-sided Gaussian (green line) was fit.  Fit values are 
($\mu$,$\sigma_l$,$\sigma_r$) = (4.33, 0.39, 0.77) for the NGC 5272 raw data, and (3.97, 0.35, 0.55) 
if the cluster was instead observed at a distance of 44.0 kpc.  Note the difference in total counts 
between the two plots; this is caused by losses due to color errors with effective distance.
}
\end{figure}

\begin{figure}
\figurenum{7}
\plotone{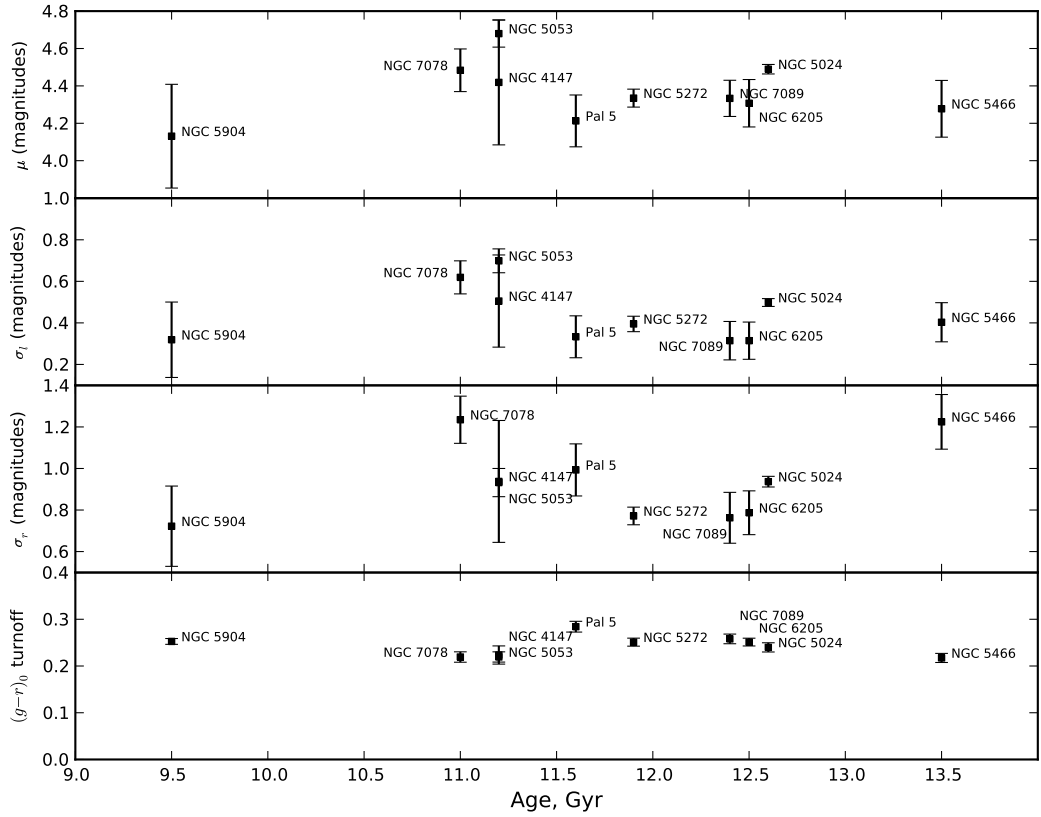}
\caption{Plot of globular cluster $\mu$, $\sigma_l$, $\sigma_r$ and turnoff color values vs. isochrone fit 
ages.  There are no statistically significant trends in our parameters as a function of cluster age.  It is 
expected that the turnoff should shift to redder color and fainter magnitude as it ages, but since these 
clusters fall along the Age-Metallicity Relationship, the change in metallicity counterbalances the effects
of the change in age.}
\label{ageparams}
\end{figure}

\begin{figure}
\figurenum{8}
\plotone{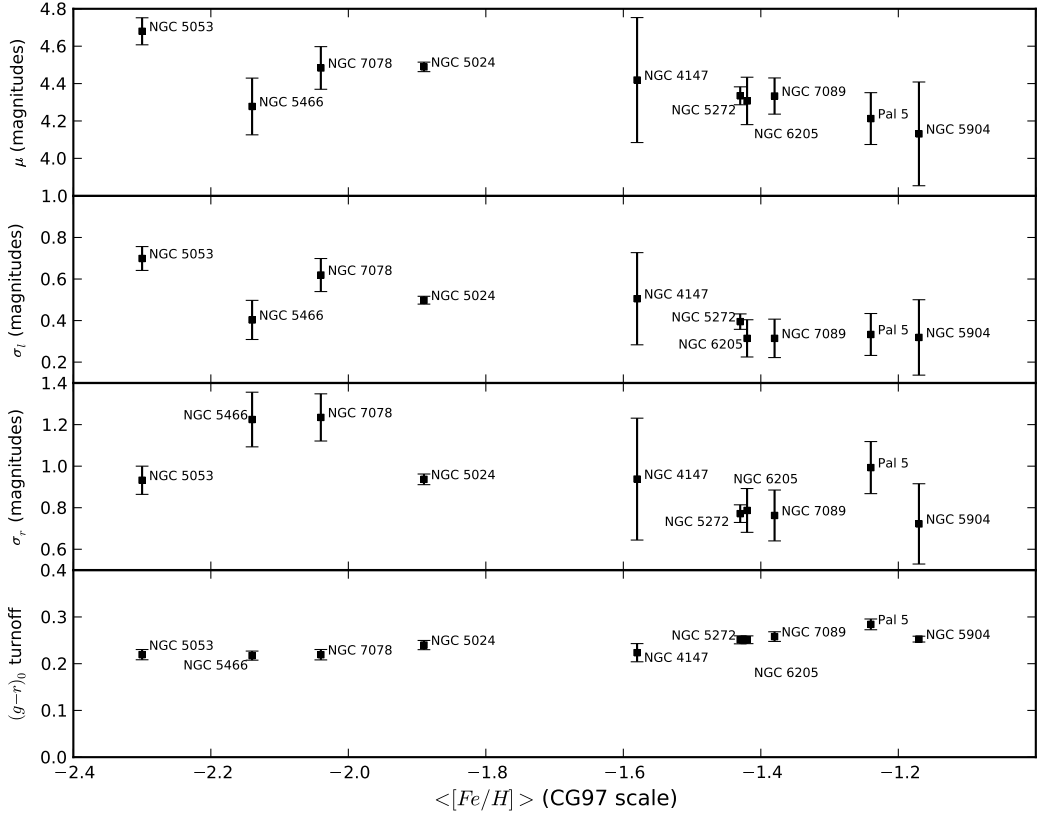}
%\epsscale{}
\caption{Plot of globular cluster $\mu$, $\sigma_l$, $\sigma_r$ and turnoff color values vs. 
CG97 scale metallicities.  The fit parameters $\mu$ and $\sigma_l$ appear to decrease with 
metallicity, indicating that isochrone fits should have slightly brighter and sharply curved 
turnoffs as $[Fe/H]$ increases.  The fit parameter $\sigma_r$ does not appear to be correlated 
with metallicity; We will show that changes in $\sigma_r$ are primarily due to distance 
(therefore, error) effects.  One would expect an increase in metallicity to produce a fainter 
turnoff, however, since the more metal-rich clusters are also younger, the turnoff is actually 
slightly brighter for a higher metallicity star.  A slight increase in the turnoff color with
metallicity is observed.}
\label{metalparams}
\end{figure}

\begin{figure}
\figurenum{9}
\plotone{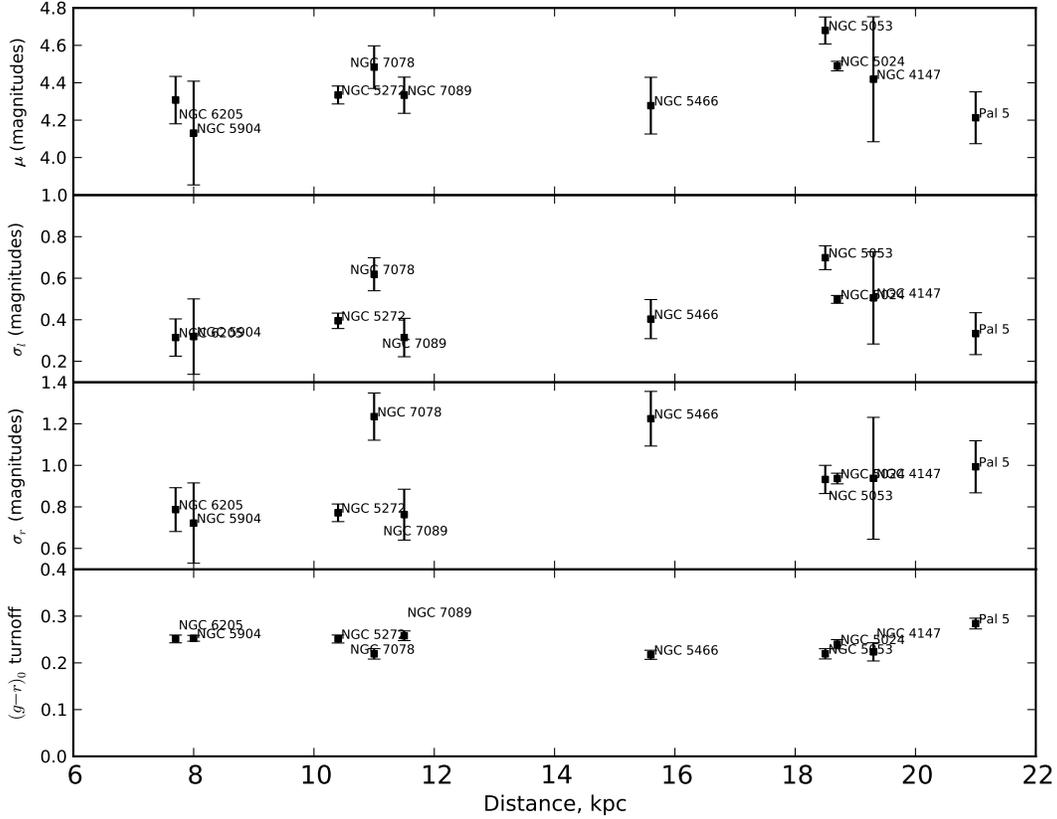}
%\epsscale{}
\caption{Plot of globular cluster $\mu$, $\sigma_l$, $\sigma_r$ and turnoff color values vs. 
cluster distances.  We see that $\mu$ appears to increase (become fainter) slightly with distance, 
$\sigma_l$ stays roughly constant, and $\sigma_r$ first increases, then decreases.  The turnoff 
color ($(g-r)_{TO}$) appears to be invariant with distance.  Since cluster properties should not 
be dependent on a cluster's distance from the Sun, any trends in this figure are a result of 
the increasing magnitude errors with distance.}
\label{distparams}
\end{figure}

\begin{figure}
\label{errorplots}
\figurenum{10}
\plotone{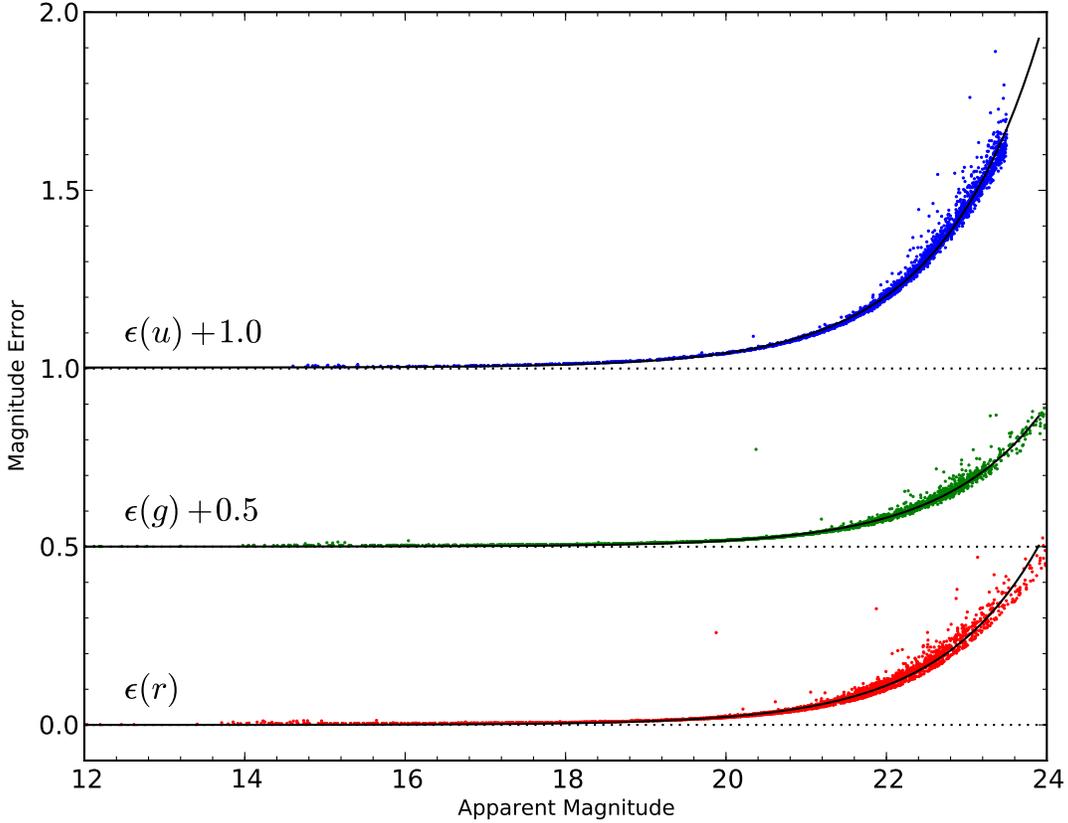}
%\epsscale{}
\caption{Plot of SDSS color magnitudes versus respective color errors.  $g_0$ and and $u_0$ errors are 
offset by constants of 0.5 and 1.0, respectively, for illustration.  Errors and dereddened photometric 
values are from the Sloan Digital Sky Survey DR7 database using stars near the globular cluster Palomar 5.  
It is clear from the plots that the errors in $u_0$ rise the fastest, followed by the errors in $r_0$.  
$u_0$ data was cut for $u_0 > 23.5$, where the $u_0$ errors became unreliable.
Black lines are functional fits to the data (colored points):
$\epsilon(u_0) =  2.71e-03 + e^{0.80 u_0 - 19.2}$; 
$\epsilon(g_0) =  3.11e-04 + e^{0.79 g_0 - 20.0}$; 
$\epsilon(r_0) = -2.63e-05 + e^{0.80 r_0 - 19.8}$. 
}
\end{figure}

\begin{figure}
\label{contam}
\figurenum{11}
\plotone{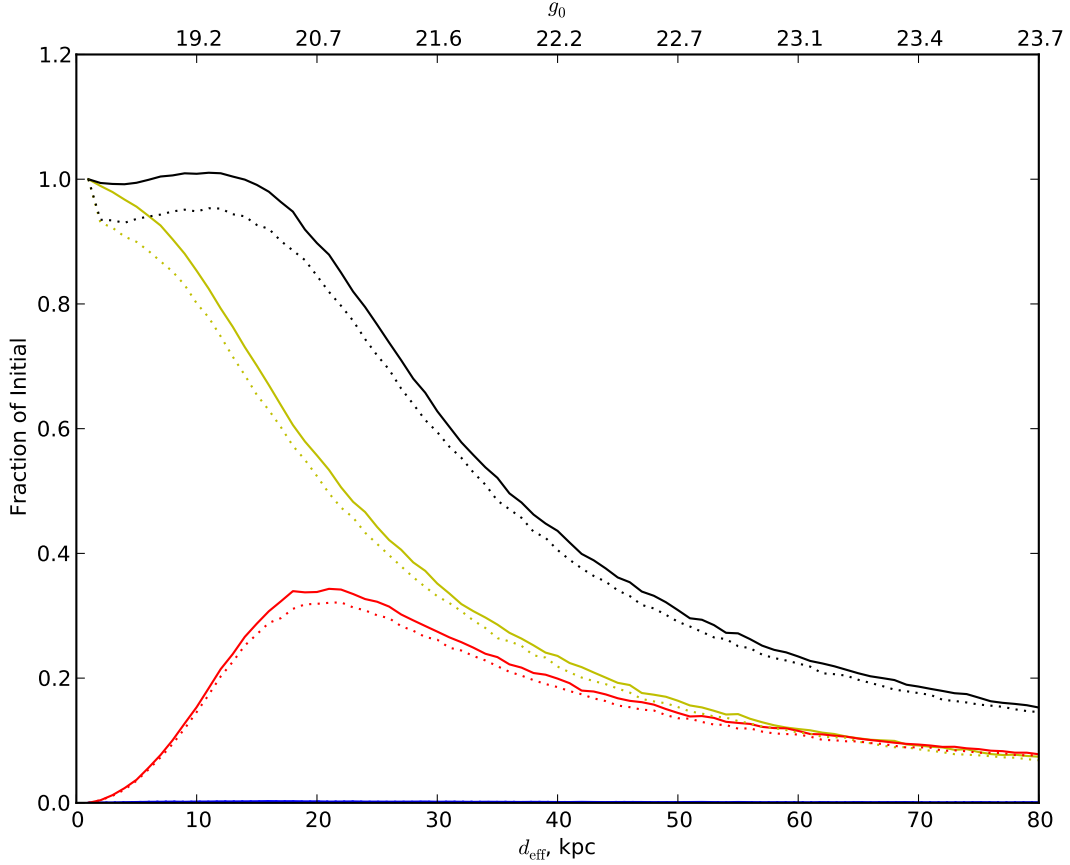}
%\epsscale{}
\caption{Percentage of stars by type detected as F turnoff stars due to SDSS color errors as a function
of distance/mean apparent magnitude.  The yellow (middle) curve indicates the percentage of `true,' (original)
 F turnoff stars detected as F turnoff stars, while the red (lower) curve indicates the percentage of `red' stars 
($(g-r)_0 > 0.3$) incorrectly identified as F turnoff stars due to color errors.  The blue curve, which is
nearly coincident with the x-axis, represents the same trend but for `blue' stars ($(g-r)_0 < 0.1$).  The 
black (top) curve is the sum of all lower curves, and illustrates the total percentage of stars detected as 
turnoff stars relative to the true turnoff distribution.  The dotted curves indicate the effect of the field 
star detection efficiency on star counts.  It is clear that the relatively higher color errors at fainter 
magnitudes, which largely consist of `red' stars,  initially causes a large number of red stars to 
contaminate the turnoff, which then tapers off as the errors increase.  True turnoff stars quickly and 
continually leak out of the turnoff bin.  By as near as 25.0 kpc from the Sun, half of all detected turnoff 
stars are actually redder star contaminants.}
\end{figure}

\begin{figure}
\label{muplot}
\figurenum{12}
\plotone{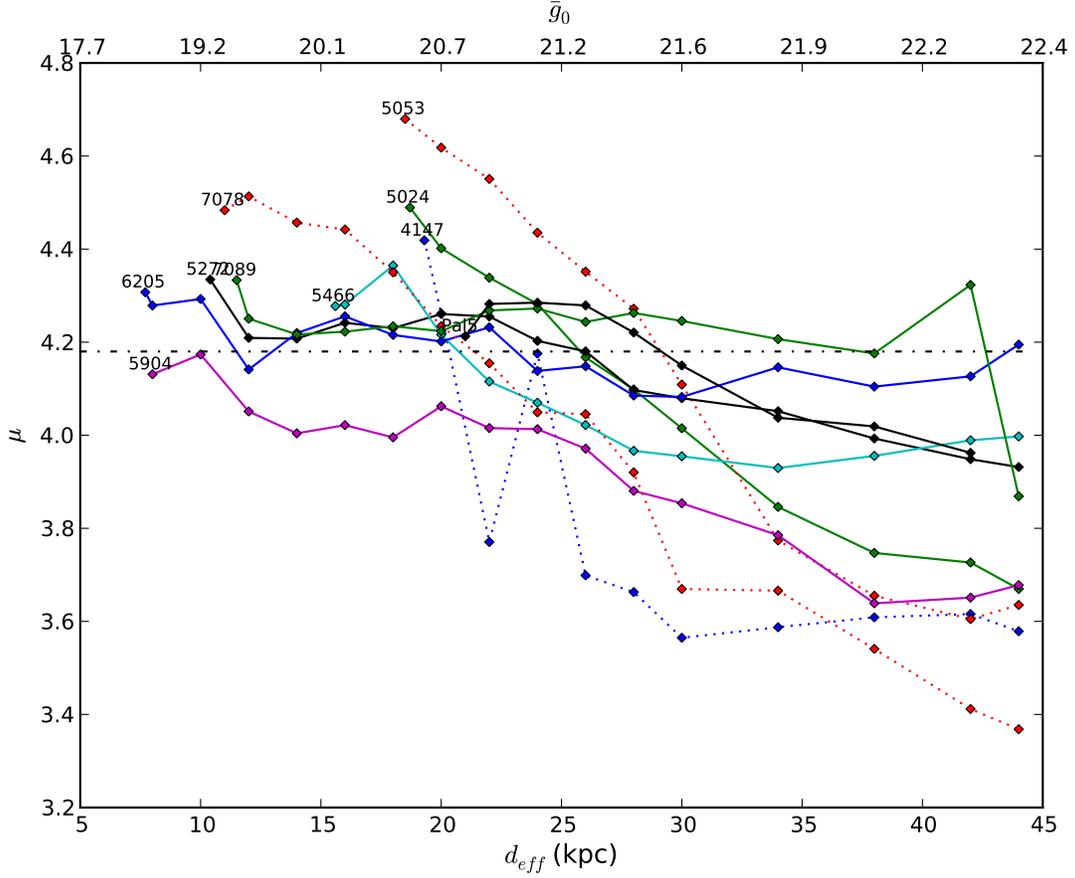}
%\epsscale{}
\caption{$\mu$ series fits to ten globular clusters, convolved to errors consistent with observing them 
at larger distances ($d_{\rm eff}$).  The cluster detection efficiency (Equation 2) was applied during the
distance shifts.  Although the fit $\mu$ values decrease with distance, the errors increase as well.  To 
within errors, the $\mu$ values are consistent with a constant value.  The red dotted series have been 
rejected as outliers.  The blue dotted series for NGC 4147 indicates large expected errors due to low star 
counts.  The error-weighted average of initial $\mu$ points for the remaining globular clusters is 
\mufit ($\pm$ \muerror), plotted as the black dot-dash line.} 
\end{figure}

\begin{figure}
\label{siglplot}
\figurenum{13}
\plotone{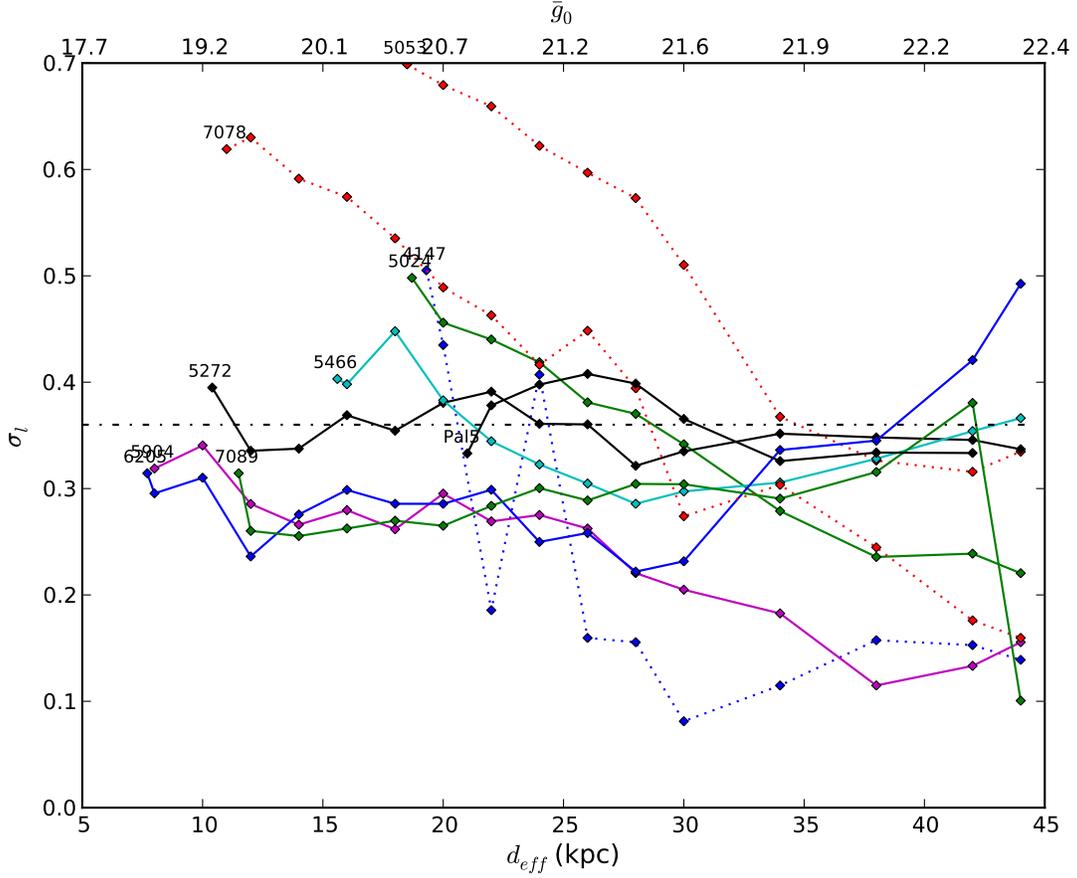}
%\epsscale{}
\caption{$\sigma_l$ series fits to ten globular clusters, convolved to errors consistent with observing them 
at larger distances ($d_{\rm eff}$).  The cluster detection efficiency (Equation 2) was applied during the
distance shifts.  To within errors, the $\sigma_l$ values stay constant.  The red dotted series 
have been rejected as outliers.  The blue dotted series for NGC 4147 indicates that large errors are expected 
due to low star counts.  The error-weighted average (ignoring outliers) of initial $\sigma_l$ points 
for the remaining clusters is \siglfit ($\pm$ \siglerror), plotted as the black dot-dash line.} 
\end{figure}

\begin{figure}
\label{sigrplots}
\figurenum{14}
\plottwo{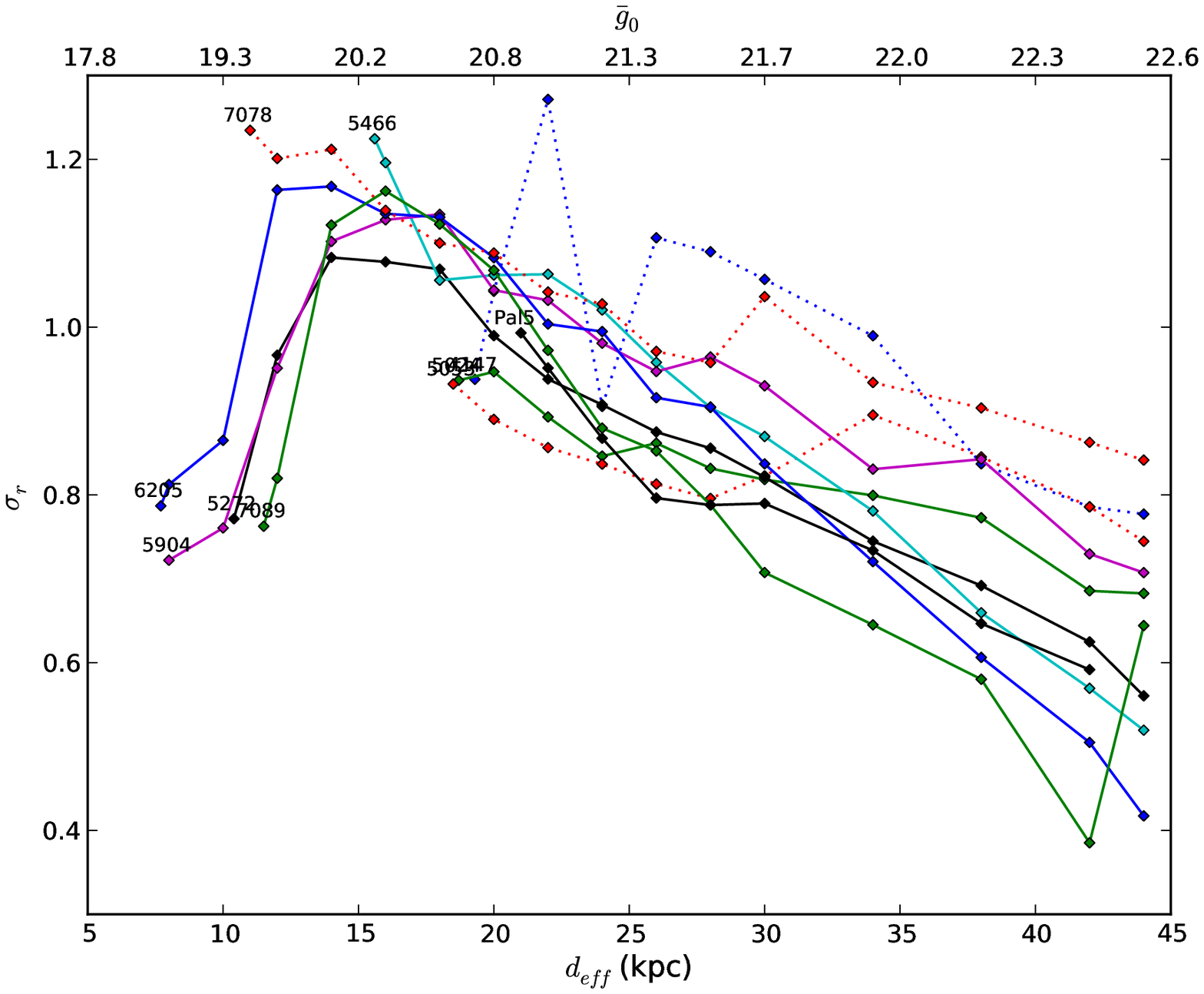}{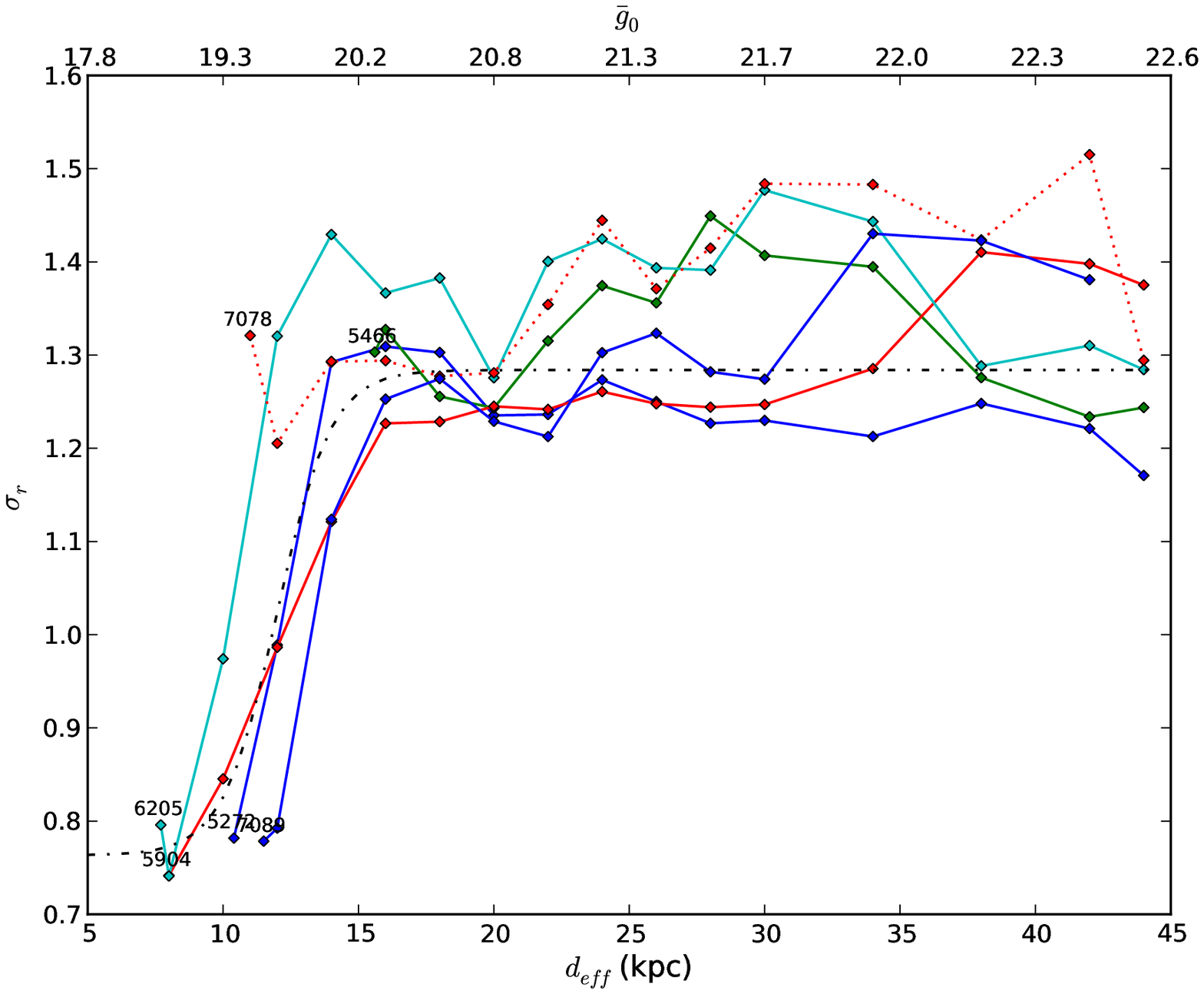}
%\epsscale{}
\caption{$\sigma_r$ series fits to ten globular clusters, convolved to errors consistent with observing them 
at larger distances ($d_{\rm eff}$).  The left plot represents clusters convolved while applying the 
SDSS crowded-field photometry detection efficiency function (Section 4).  When this detection efficiency 
is applied, all clusters in our analysis follow the same trend (reproducing the observed dependence of 
$\sigma_r$ with distance) as they are convolved to farther $d_{\rm eff}$.  The right plot contains only 
clusters not initially affected by the detection efficiency, and did not have it applied during the convolutions, 
since it was not possible to remove the crowding effect from the more distance cluster data. This represents the 
ideal observation, in which no SDSS detection losses occur beyond the initial status of the cluster, but with 
only photometric errors increasing with distance.  In both cases a quick rise in $\sigma_r$ with distance is 
observed, due to the large influx of redder main-sequence stars into the F turnoff cut, as described in Figure 5.  
In the left plot,  the detection efficiency continually cuts into the fainter stars, causing a constant fall 
in $\sigma_r$ after the quick rise.  When there is no loss due to detection efficiency (right), $\sigma_r$ 
remains constant after the initial rise.  To this trend we fit a sigmoid function (dashed line, see text):  
$\alpha$, $\beta$, and $\gamma$ values of \alphafit ($\pm$ \alphaerror), \betafit ($\pm$ \betaerror), and 
\gammafit ($\pm$ \gammaerror), respectively.  Dotted lines indicate clusters that have been treated as outliers.}
\end{figure}

\begin{figure}
\figurenum{15}
\plotone{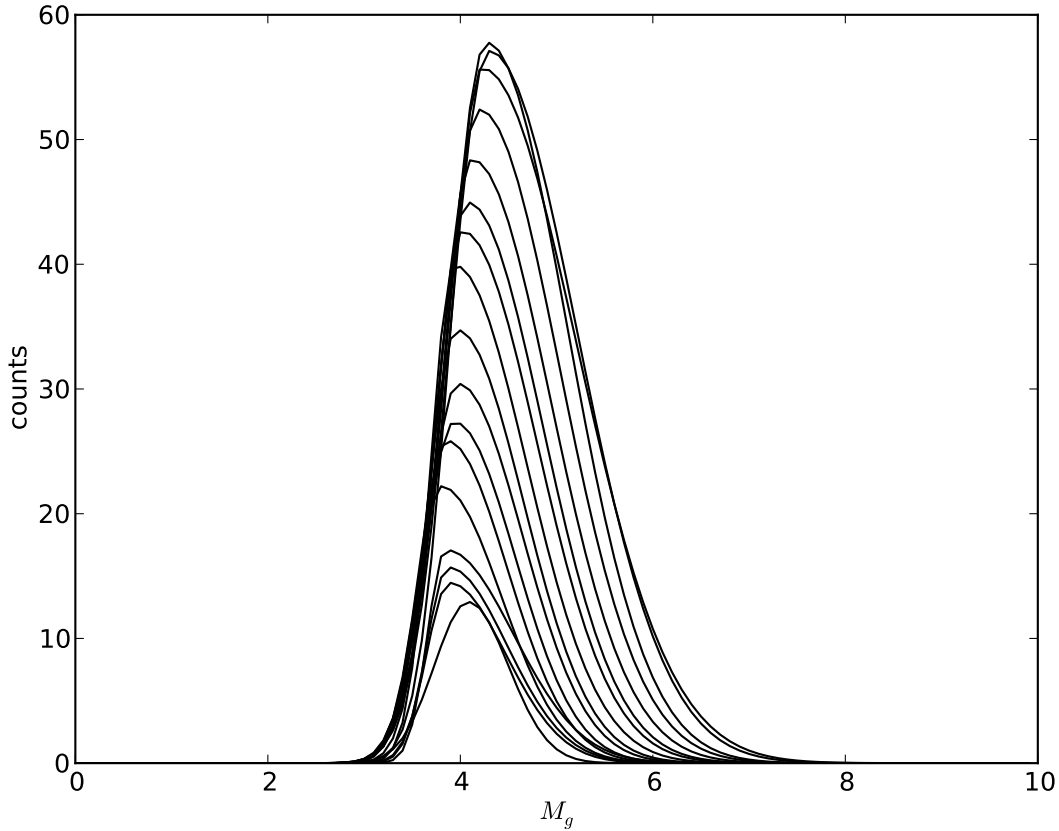}
%\epsscale{}
\caption{Double-sided Gaussian fits to NGC 6205 turnoff ($0.1 < (g-r)_0 < 0.3$) histograms with increasing 
effective distance ($d_{\rm eff}$).  The third tallest Gaussian is at the original distance of 7.7 kpc, while 
the remaining Gaussians are fits to the data shifted to $d_{\rm eff}$s of 8.0 kpc through 28.0 kpc, in 2.0 kpc
increments.  Initially, the Gaussian fits become taller and $\sigma_r$ becomes larger as redder contamination 
stars enter the data set.  Then the peak of the histogram becomes smaller as the cluster moves to larger 
$d_{\rm eff}$ values, where increasing color errors cause a bleed-off of turnoff stars.  Note that as the cluster 
is shifted beyond 20.0 kpc (fourth curve from the bottom), the $\sigma_r$ value eventually begins to shrink as 
the cluster detection efficiency removes fainter stars from the data.}
\label{gaussseries}
\end{figure}

\begin{figure}
\figurenum{16}
\plotone{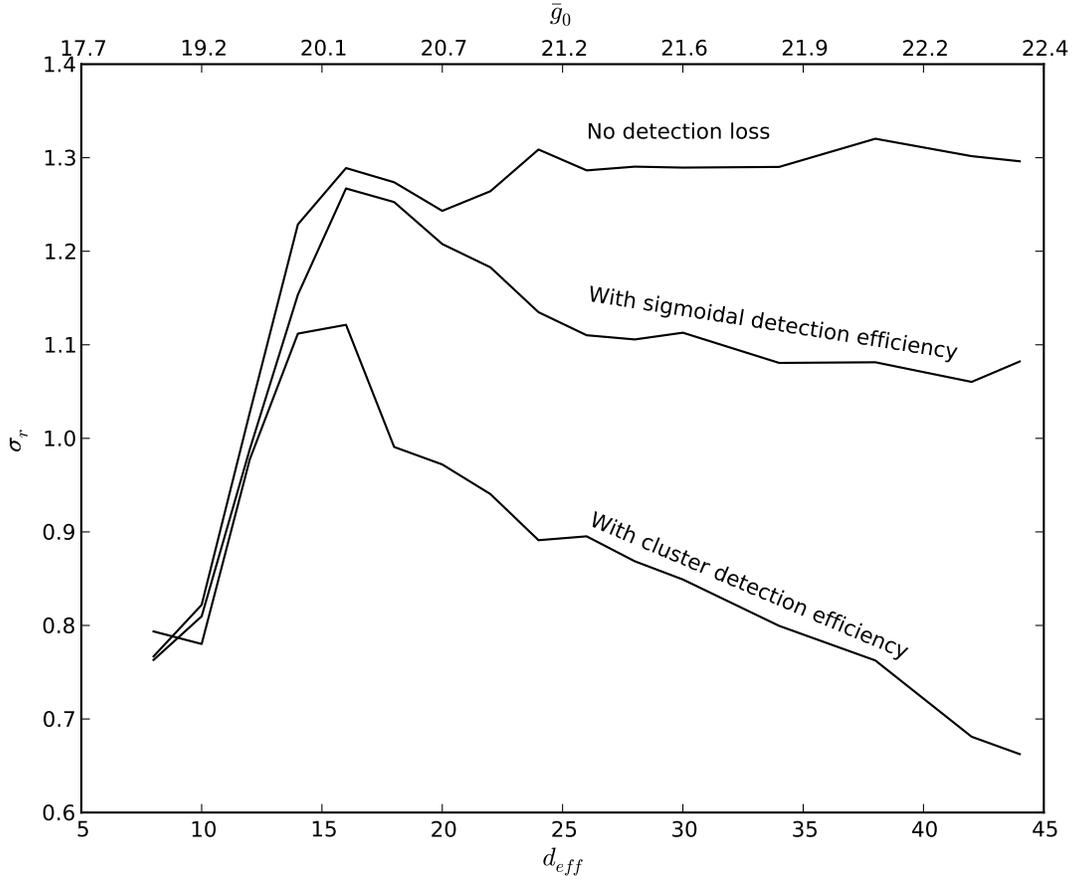}
%\epsscale{}
\caption{Weighted averages of $\sigma_r$ series fits with effective distance ($d_{\rm eff}$) for three different
detection efficiency functions:  no detection efficiency applied, (100\% detection, upper curve) the sigmoidal
SDSS detection efficiency from \citet{nyetal02}, (middle curve) and the SDSS crowded-field photometry parabolic
cluster detection efficiency discussed in Section 4 (lower curve).}
\label{sigrcompare}
\end{figure}

\begin{figure}
\figurenum{17}
\plotone{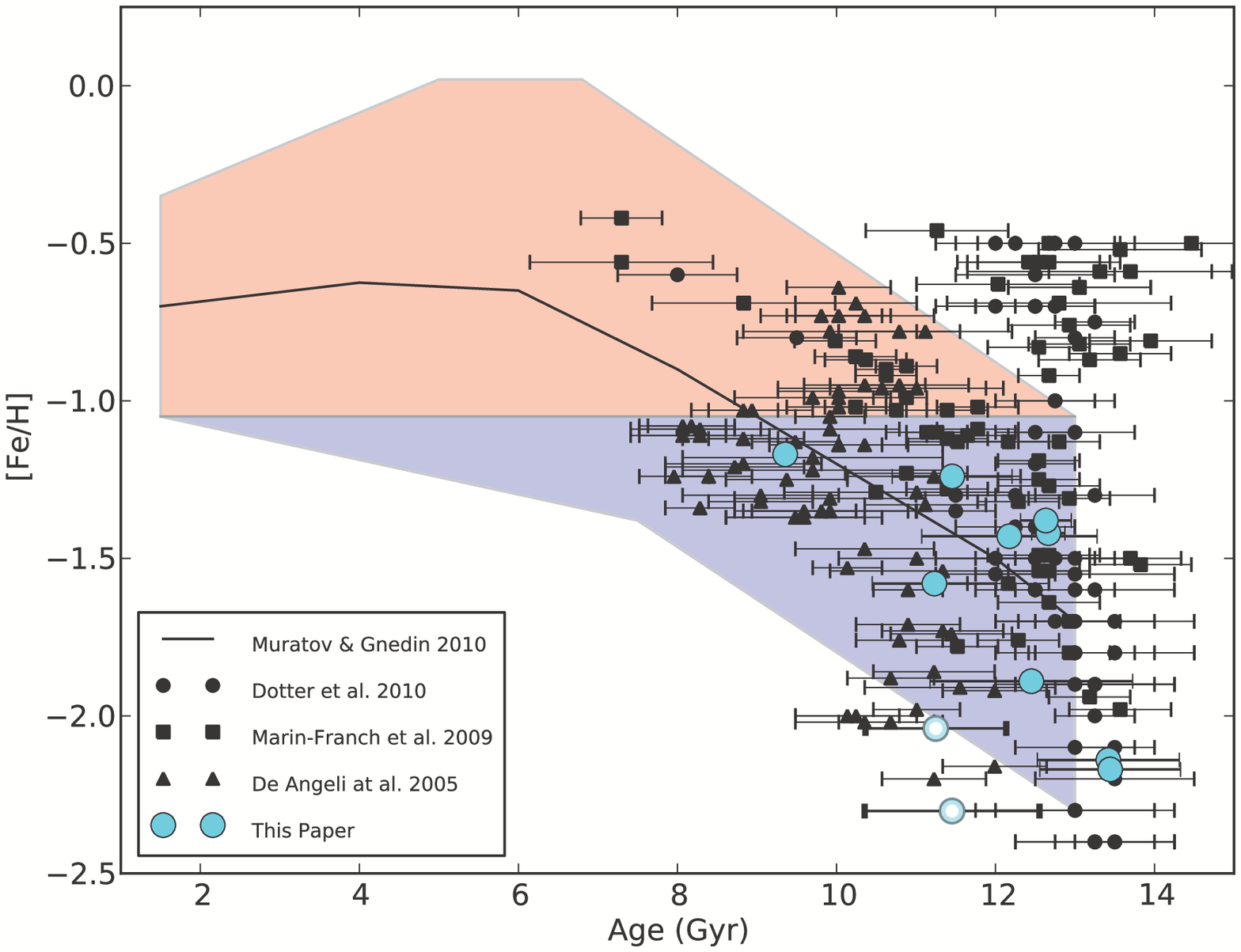} 
%\epsscale{}
\caption{Plot of globular cluster ages versus metallicity, from four different sources: this paper 
(blue circles), \citet{D2010, DA2005, MF2009}.  Two clusters, NGC 5053 and NGC 7078, are shown as blue 
rings to indicate that they are outliers in our analysis.  These are overlaid on a theoretical 
age-metallicity relationship from \citet{mg10} (blue- and red-shaded areas).  Note that most of the clusters, 
including all of the clusters in this study besides noted outliers NGC 5053 and NGC 7078, are consistent with 
the theoretical age-metallicity relationship.  Note that there are some high age, high metallicity clusters 
in the \citet{MF2009} and \citet{D2010} data that they attribute to a constant age with metallicity 
relationship at around 13 Gyr in the inner halo.}
\label{amriso}
\end{figure}

\begin{figure}
\figurenum{18}
\plotone{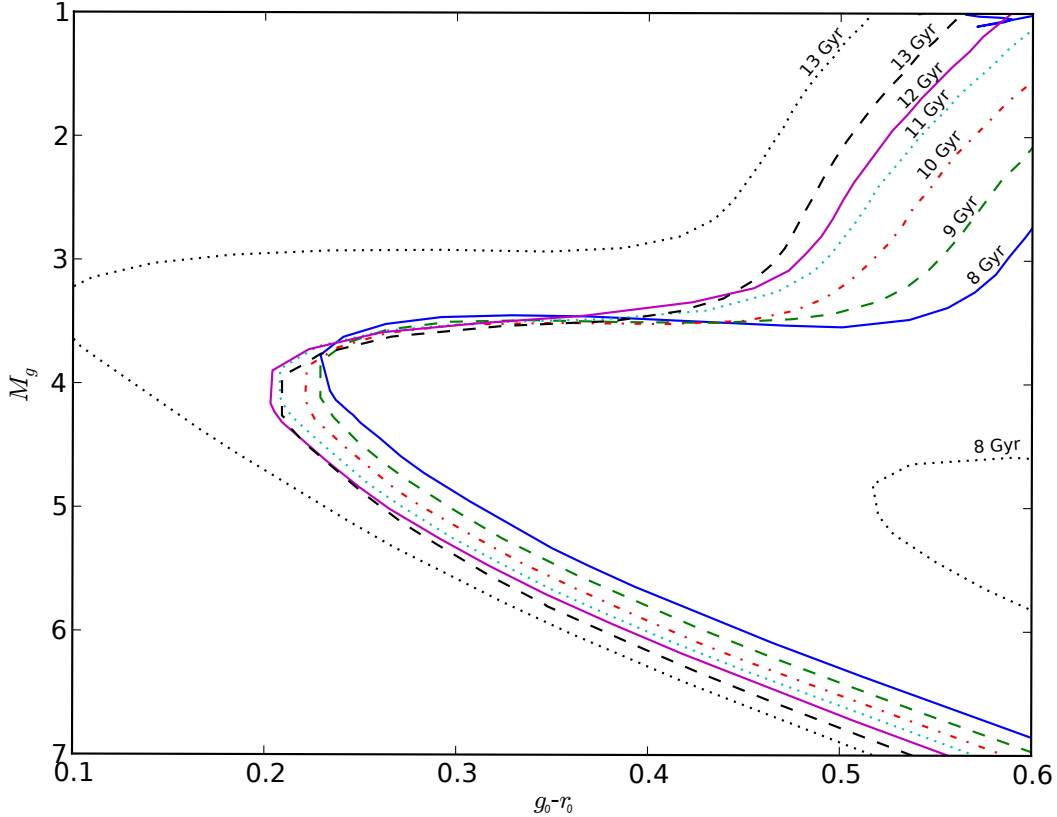}
%\epsscale{}
\caption{Plot of Padova isochrones, modified by a correctional fit (Equation 1), from the Age-Metallicity 
Relationship (AMR) presented in \citet{mg10}. Two isochrones not on the AMR are also plotted as black dotted 
series, at 8 Gyr and 13 Gyr, in order to show the behavior of extreme outliers.  All of the isochrones on the 
AMR have turnoff values that are similar to each other, with a turnoff color between between $0.2 < (g-r)_0 < 0.23$ 
and a turnoff magnitude of $3.77 < M_g < 4.16$.  The subgiant branch is also well constrained in magnitude:
$3.4 < M_g < 3.6$.  These constraints will be useful in determining distances to old stellar populations.  The 
following Age (Gyr) and [Fe/H] (dex) value sets were used to produce the AMR isochrones:  13.0, -1.7; 12.0, 
-1.5; 11.0, -1.35; 10.0, -1.20; 9.0, -1.05; 8.0, -0.90.  The age, [Fe/H] values for outlying isochrones are: 
13.0, 0.0; 8.0, -2.0}
\label{amr}
\end{figure}

\end{document}